\documentclass[aps,pra,showpacs,twocolumn,groupedaddress]{revtex4}
\usepackage{hyperref}
\usepackage{graphicx}
\usepackage{amsmath}
\usepackage{amsfonts}
\usepackage{amssymb}
\usepackage{bm}
\usepackage{xcolor}

\begin{document}

\title{Phase separation of binary condensates in harmonic and lattice 
       potentials}

\author{S. Gautam and D. Angom}
\affiliation{Physical Research Laboratory,
         Navarangpura, Ahmedabad - 380 009\\}

\date{\today}

\begin{abstract}
We propose a modified Gaussian ansatz to study binary condensates, trapped in 
harmonic and optical lattice potentials, both in miscible and immiscible 
domains. The ansatz is an apt one as it leads to the smooth transition from 
miscible to immiscible domains without any {\em a priori} assumptions.
In optical lattice potentials, we analyze the squeezing of the density
profiles due to the increase in the depth of the optical lattice potential.
For this we develop a model with three potential wells, and define the
relationship between the lattice depth and profile of the condensate.
\end{abstract}



\maketitle


\section{\label{section-1}Introduction}
After the successful experimental realization of two-species
Bose-Einstein condensate (TBEC), consisting of different hyperfine spin 
states of $^{87}$Rb \cite {Myatt,Mertes,Thalhammer}, experimental as well as 
theoretical investigations in this field have come a long way. TBECs of 
different atomic species ($^{41}$K and $^{87}$Rb) \cite {Modugno} and of 
different isotopes of same atomic species ($^{85}$Rb and $^{87}$Rb) \cite 
{Papp} have been experimentally observed. Phase separation, a typical feature 
of two component Bose-Einstein condensates, has been observed unambiguously 
\cite{Papp}. A lot of static as well as dynamical properties of the TBECs 
have been analyzed in great detail in recent years. These include the ground 
state geometry \cite{Ho,Chui1,Chui2, Trippenbach,Timmermans,Ao,Barankov,
Schaeybroeck,Gautam1}, modulational instability \cite{Kasamatsu1,Raju,Ronen}, 
Rayleigh-Taylor instability \cite{Gautam2},\cite{Sasaki}, Kelvin-Helmholtz 
instability \cite{Takeuchi}, etc..

 The ground state geometry of the TBEC has been studied, semi-analytically,
using Thomas-Fermi (TF) approximation \cite{Ho,Chui1,Chui2,Trippenbach}.
These works did not take into account the contribution of the interface
energy explicitly. The interface energy was taken into account in some
other works \cite{Timmermans,Ao,Barankov}; however, the interface energy
correction incorporated in these works was not good enough to reproduce the
experimentally observed ground state structures with planar interface
\cite{Papp}. A more accurate analytic approximation to account for
interface energy was suggested by Schaeybroeck et al. \cite{Schaeybroeck},
and a recent work \cite{Gautam1} conclusively proved that using this
analytic approximation for interface energy in TF regime, planar and
cylindrical geometries emerge as the ground state structures in cigar and
pan-cake shaped trap potentials. The common salient feature of all
these works is the use of TF approximation, which is a good approximation
for large number of atoms of each species. In fact if $N,~a$, and
$a_{\rm osc}$ are respectively, number of atoms of the component species,
s-wave scattering length, and oscillator length, TF approximation is valid
provided $Na/a_{\rm osc}\gg1$. Obviously, this condition is not satisfied
for attractive condensates. For small number of atoms as well, say of the 
order of a few hundreds, and for very weakly interacting condensates
($a/a_{\rm osc}\ll1$), the contribution of the kinetic energy to the total
energy is significant and can not be merely treated as a correction to total
energy as is done in TF based approaches. For such TBECs, TF approximation
is not a good approximation, and hence can not be relied upon to determine
the ground state structure of the TBEC. In present work, we analyze the
ground sate properties of the TBECs using suitable ansatz in both miscible
and immiscible regimes. The Gaussian nature of the ansatz makes our approach
better equipped to study the stationary state properties of the of very
weakly interacting ($Na/a_{\rm osc}\sim 1$) TBECs and also those with
attractive interactions. With the advent of Feshbach resonances \cite{Chin}, 
it is experimentally possible to tune scattering lengths to reach the very 
weakly interacting regime or even non-interacting regime \cite{Weber, 
Kraemer, Roati}. Magnetic Feshbach resonances can tune only one scattering 
length independently, whereas optical Feshbach resonances \cite{Fedichev} 
open up the possibility of tuning different scattering length
in a multicomponent system independently. With the experimental observation
of optical Feshbach resonances in $^{172}$Yb\cite{Enomoto} and $^{174}$Yb 
\cite{Yamazaki}, experimental realization of weakly interacting regime in 
binary condensates appears a distinct possibility. The present approach, 
thus, supplements TF based semi-analytic schemes to determine the ground 
state geometries of the TBECs.

Keeping a pace with the studies on TBECs, has been the realization of various
condensed matter phenomena like ac Josephson effect, Bloch oscillations,
Landau-Zener tunneling, etc. in single species Bose-Einstein condensates
(BECs) trapped in optical lattices \cite{Anderson,Burger,Cataliotti,Morsch}.
Superfluid-Mott insulator transition has been also observed with BECs in
optical lattice potentials \cite{Greiner}. In mean field approximation, we
also study the ground sate geometry of TBECs trapped in optical lattice
potentials using  discrete nonlinear Schr\"odinger equation (DNLSE)
\cite{Trombettoni}. We also compare our semi-analytic results with the
numerical solution of coupled Gross-Pitaevskii equations and find a very
good agreement especially in very weakly interacting regime.


\section{TBECs in axisymmetric traps}
In this section, we provide a general variational scheme to study the 
stationary state geometry of TBECs in axisymmetric trap potentials
\begin{equation}
  V_i(r,z) = \frac{m_i\omega^2}{2}(r^2 + \alpha_i^2 z^2),
\label{eq.pots}
\end{equation}
where $i=1,2$ is the species index, $\omega$ is the radial trap frequency
for two components, and $\alpha_i$ are the anisotropy parameters. For 
simplicity of analysis, we consider trap potentials for the component
species are identical $\alpha_1=\alpha_2=\alpha$ and $m_1=m_2=m$. The ground
state of the TBEC is described by a set of coupled GP equations
\begin{equation}
  \left[ \frac{-\hbar^2}{2m}\nabla^2 + V_i(r,z) + 
  \sum_{j=1}^2U_{ij}|\Psi_j|^2 \right]\Psi_i = \mu_i\Psi_i, 
\label{eq.gp}
\end{equation}
in mean field approximation, where $i = 1, 2$ is the species index. Here 
$U_{ii} = 4\pi\hbar^2a_{ii}/m_i$, where $m_i$ is the mass and $a_{ii}$ is 
the $s$-wave scattering length, is the intra-species interaction, 
$U_{ij}=2\pi\hbar^2a_{ij}/m_{ij}$, where $m_{ij}=m_i m_j/(m_i+m_j)$ is the
reduced mass and $a_{ij}$ is the inter-species scattering length, is the 
inter-species interaction, and $\mu_i$ is the chemical potential of the 
$i$th species. The energy of the TBEC is
\begin{eqnarray}
E & = & \int_{-\infty}^{\infty}\left[\sum_{i=1}^2 \left(\frac{\hbar^2}{2m}
        |\nabla\Psi_i|^2+ V_i(r,z)\Psi_i^2  \right. \right . 
             \nonumber \\ 
  &   &\left . \left . + \frac{U_{ii}}{2}|\Psi_i|^4\right)
       + U_{12}|\Psi_1|^2|\Psi_2|^2\right]d{\bf r}.
\label{e_axisym}			 
\end{eqnarray}
To rewrite the energy in suitable units, define the oscillator length of the 
trapping potential
\begin{equation}
   a_{\rm osc} = \sqrt{\frac{\hbar}{m\omega}},
\end{equation}
and consider $\hbar\omega$ as the unit of energy. We then divide the 
Eq.(\ref{e_axisym}) by $\hbar\omega$ and apply the transformations
\begin{equation}
\tilde{r}   =  \frac{r}{a_{\rm osc}}, ~\tilde{z}  = \frac{z}{a_{\rm osc}},
    ~\tilde{t} =  t\omega, \text{and} ~\tilde{E}= \frac{E}{\hbar\omega}.
\end{equation}
The transformed order parameter  
\begin{equation}
   \phi_{i}(\tilde{r},\tilde{z})=   \sqrt{\frac{a_{\rm osc}^3}{N_i}}
       \Psi_i(r,z), 
\label{scaled_wavefun}
\end{equation}
and energy of the TBEC in scaled units is
\begin{eqnarray}
\tilde{E} & = & \int d\tilde{{\bf r}}\left\{\sum_{i=1}^2 N_i\left[\frac{1}{2}
       |\nabla \phi_i|^2+ V_i(\tilde{r},\tilde{z}) |\phi_i|^2 +     
                          \right.\right.\nonumber \\
  &   &\left. \left.N_i\frac{\tilde{U}_{ii}}{2}|\phi_i|^4 \right] + 
        N_1 N_2 \tilde{U}_{12} |\phi_1|^2|\phi_2|^2\right\},
\label{escal_axisym}			 
\end{eqnarray}
where $\tilde{U}_{ii} = 4\pi a_{ii}/a_{\text{osc}}$ and 
$\tilde{U}_{12}=4 \pi  a_{12}/a_{\text{osc}}$  in the scaled units. For 
simplicity of notations, from here on we represent the transformed quantities 
without tilde. We obtain the coupled 2D GP equations, 
\begin{equation}
  \left[ -\frac{1}{2}\frac{\partial ^2}{\partial r ^2} 
        -\frac{1}{2r}\frac{\partial}{\partial r} + V_i(r,z) + 
  \sum_{j=1}^2 G_{ij}|\phi_j|^2 \right]\phi_{i} = \mu_i\phi_{i}, 
\label{2d.gp}
\end{equation}
when the energy functional ${\cal E} = E - \sum_i \mu_i N_i$ is variationally 
minimized with $\phi_i^\star$ as parameters of variation, here $G_{ii} = 
N_i \tilde{U}_{ii}$ and $G_{ij} = N_j\tilde{U}_{ij}$. The equations are then 
solved numerically. Another approach ideal for semi-analytic treatment is to 
adopt a predefined form of $\phi$ with few variational parameters and 
minimize $E$. This is outlined for the 2D and 1D in the next subsections.


\subsection{Variational ansatz}

As mentioned earlier, the ground state of TBEC can be either be miscible or
immiscible depending on the interaction parameters. An ansatz which describes 
the ground state of the TBEC well, both in the miscible and immiscible domain, 
when $a_{ii}, a_{ij} > 0$ and $a_{22} > a_{11}$ is
\begin{eqnarray}
  \Psi_1(r,z) & = & a e^{-(r^2+\alpha^2z^2)/(2b^2)},\nonumber\\
  \Psi_2(r,z) & = & \left[f + c (r^2+\alpha^2z^2)\right ] e^{-(r^2
                    + \alpha^2 z^2)/(2d^2)},
  \label{var_ansatz_3D}
\end{eqnarray}
with $a,~b,~ f,~c$, and $d$ as the variational parameters. In the phase 
separated or immiscible domain, this ansatz is apt for the ellipsoidal
interface geometry where the density distribution follows the equipotential
surfaces of the trapping potentials. It is applicable, in particular, to weakly 
interacting TBECs, and this is precisely the underlying assumption for 
choosing the present ansatz.  This ansatz is not suitable for analysis 
of planar and cylindrical geometries where density distributions do not follow 
equipotential surfaces. A similar ansatz, perhaps more general, is used in 
ref. \cite{Adhikari} to examine symbiotic gap and semi gap solitons in TBECs 
trapped in optical lattice potentials.  The parameter $c$ accounts for the 
flattening of the density profile of the second species as the intra-species 
non-linearity is increased. Moreover, for symmetric ground state geometries, 
$c$ is a parameter related to the overlap of two component wave functions. For 
an ideal case when all the non-linearities are small and equal, $c$ is 
$\approx0$, and hence the two species of the TBEC completely overlap. 

  If $N_1$ and $N_2$ are the number of particles of the two species, then 
from Eq.(\ref{scaled_wavefun}) in scaled units 
\begin{equation}
  \int_{-\infty}^{\infty}d{\bf r} |\phi_i(r,z)|^2 = 1.
  \label{normalization_3D}
\end{equation}
Since the number of atoms of each species is fixed, the normalization
conditions are equivalent to two constraint equations and reduce the number 
of variational parameters by two. After evaluating the integrals, the two 
constraint equations are
\begin{eqnarray}
b & = & \frac{(-1)^{2/3} \alpha ^{1/3}}{a^{2/3} \sqrt{\pi }}, \nonumber\\
c & = & \frac{2\sqrt{3}}{15 d^7}\left(\frac{\sqrt{ 5d^7\pi^{3/2}\alpha -
        2 d^{10} f^2 \pi ^3}}{\pi ^{3/2}} - \sqrt{3} d^5f \right ) .
\end{eqnarray}
From Eqs.(\ref{escal_axisym}-\ref{normalization_3D}), the energy of the 
first species is 
\begin{equation}
   E_1  =  \frac{a^2 b N_1\pi ^{3/2}}{8 \alpha} \left (
           \sqrt{2} a^2 b^2 G_{11}+ 4+6 b^4+2\alpha ^2 \right),
\end{equation}
similarly, for the second species
\begin{eqnarray}
  E_2 & = & \frac{dN_2\pi ^{3/2}}{2048 \alpha}\left \{ 128 \left[4 f^2 
            \left(2+3 d^4+\alpha ^2\right) + 4 c d^2 f ( 2 
                 \right.\right.  \nonumber\\
      &   & \left. + 15d^4 + \alpha ^2) + c^2 d^4 (22 + 105 d^4 + 11 \alpha ^2)
            \right] \nonumber \\
      &   &  +\sqrt{2} d^2 G_{22}(945 c^4 d^8+1680 c^3 d^6 f+ 1440 c^2 d^4 f^2
                   \nonumber \\
      &   &  \left. +768 c d^2 f^3+256 f^4)\right\},
\end{eqnarray}
and the energy from the inter species interaction is
\begin{eqnarray}
 E_{12} & = & \frac{2N_1G_{12}b^2 d^3\pi ^{3/2}}{2048\alpha(b^2+d^2)^{7/2}}
              \left[15 b^4 c^2 d^4 + 12b^2cd^2f (b^2 \right. 
                     \nonumber \\
        &   &  \left. +d^2) + 4f^2(b^2+d^2)^2 \right].
\end{eqnarray}
We can then define the energy per boson as
\begin{equation}
   \epsilon = \frac{E_1 + E_2 + E_{12}}{N_1+N_2}.
\end{equation}

\begin{center}
\begin{figure}
\begin{tabular}{cc}
\resizebox{40mm}{!}{\includegraphics[angle=-90]{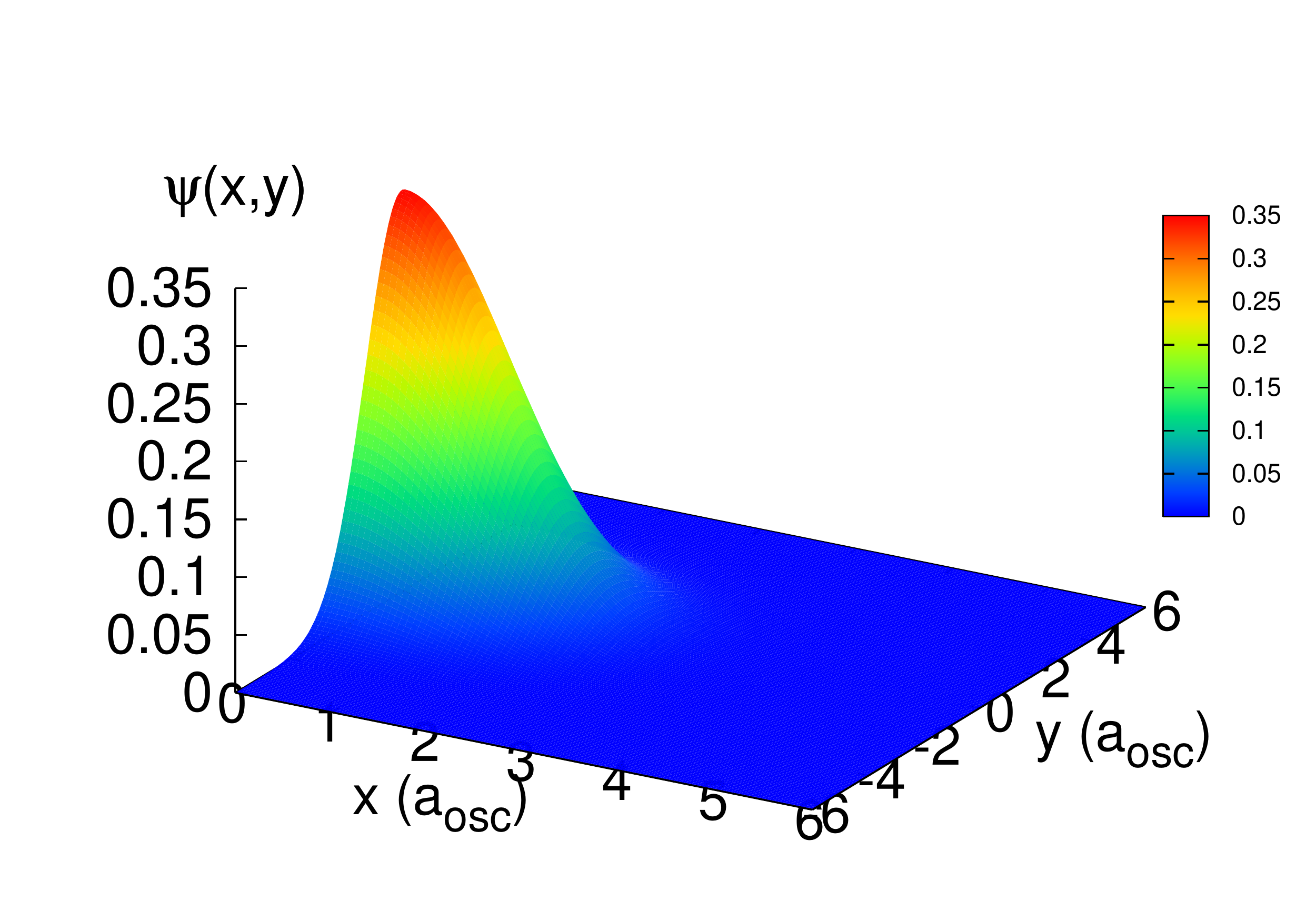}} &
\resizebox{40mm}{!}{\includegraphics[angle=-90]{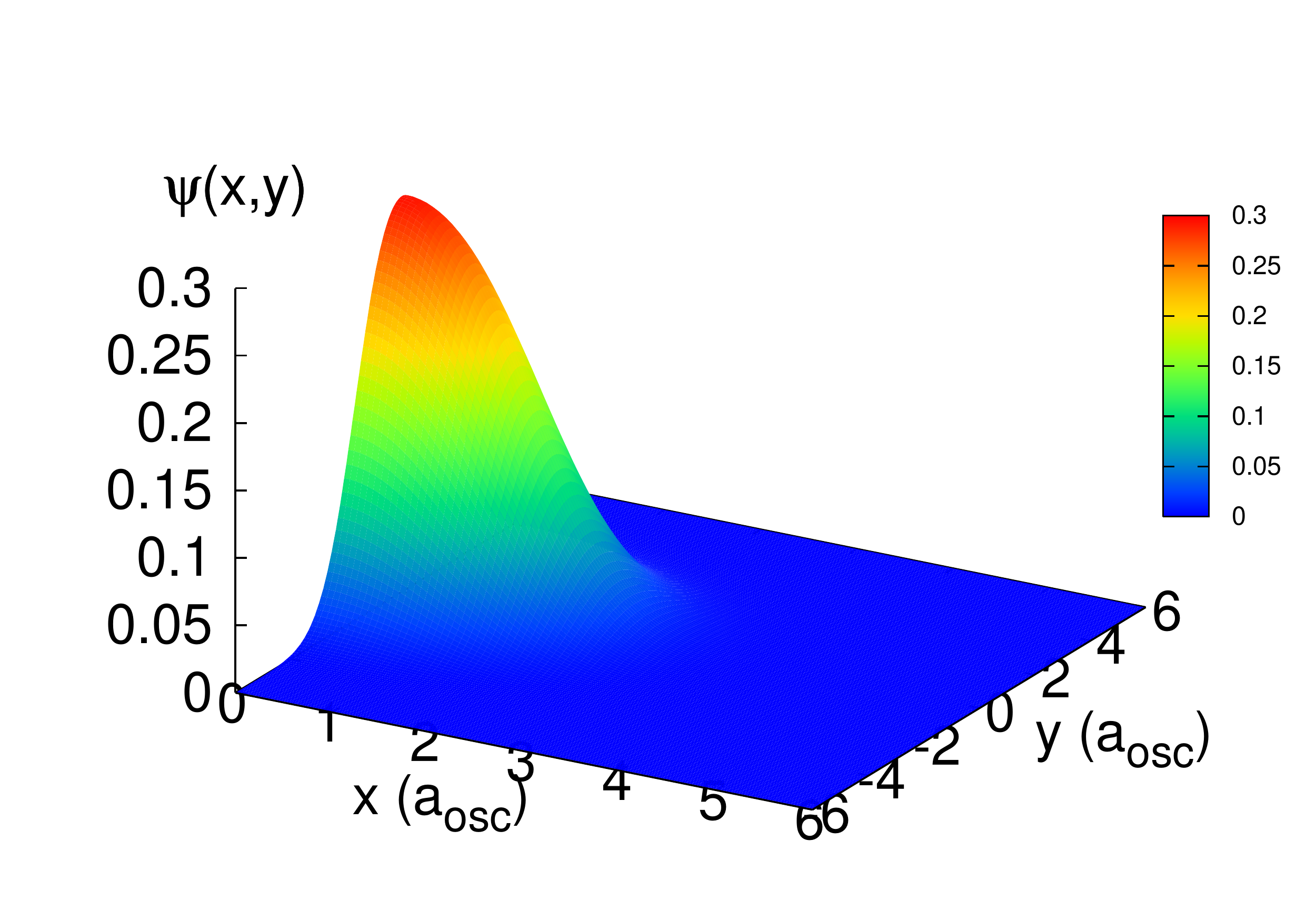}} \\
\resizebox{40mm}{!}{\includegraphics[angle=-90]{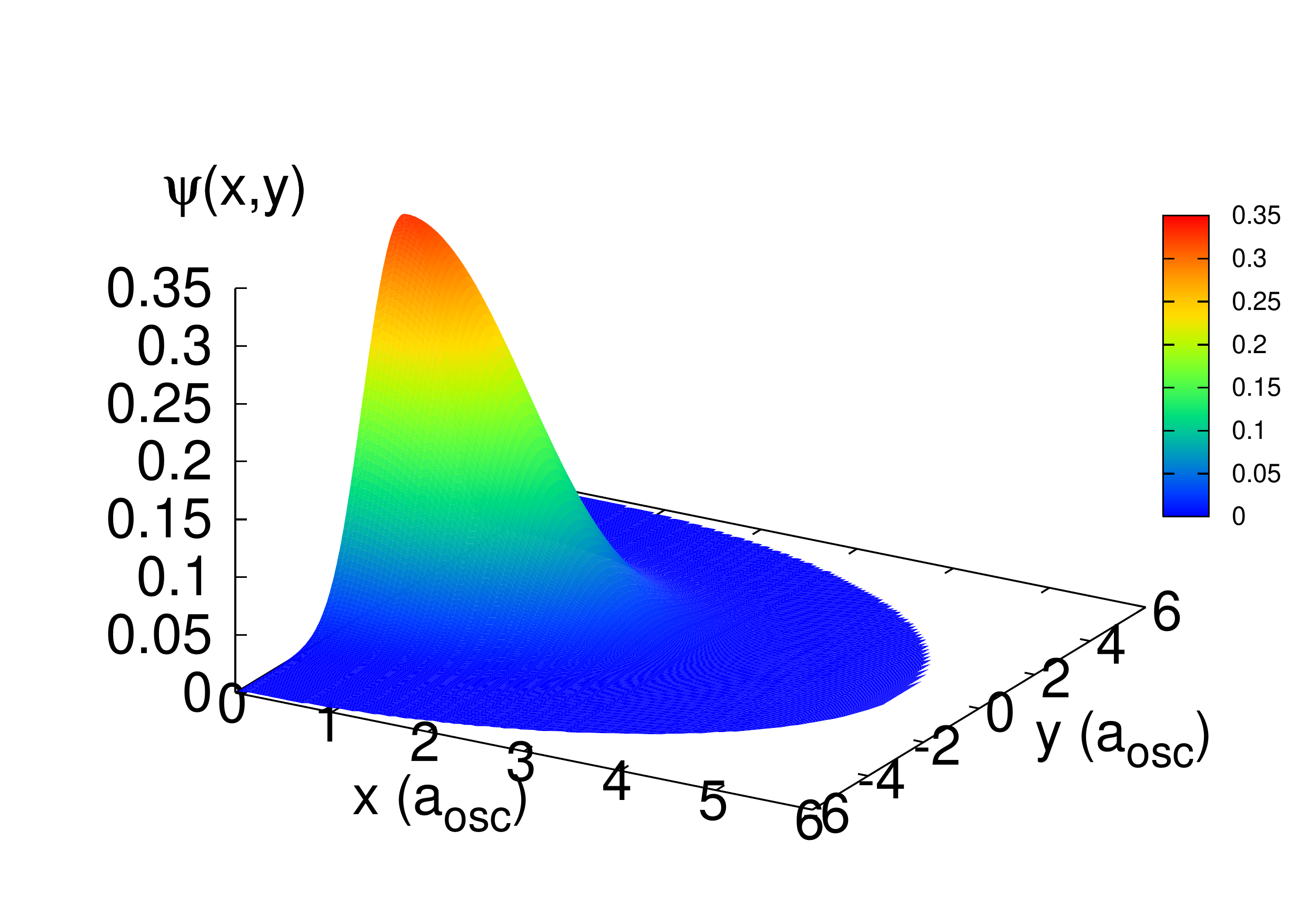}} &
\resizebox{40mm}{!}{\includegraphics[angle=-90]{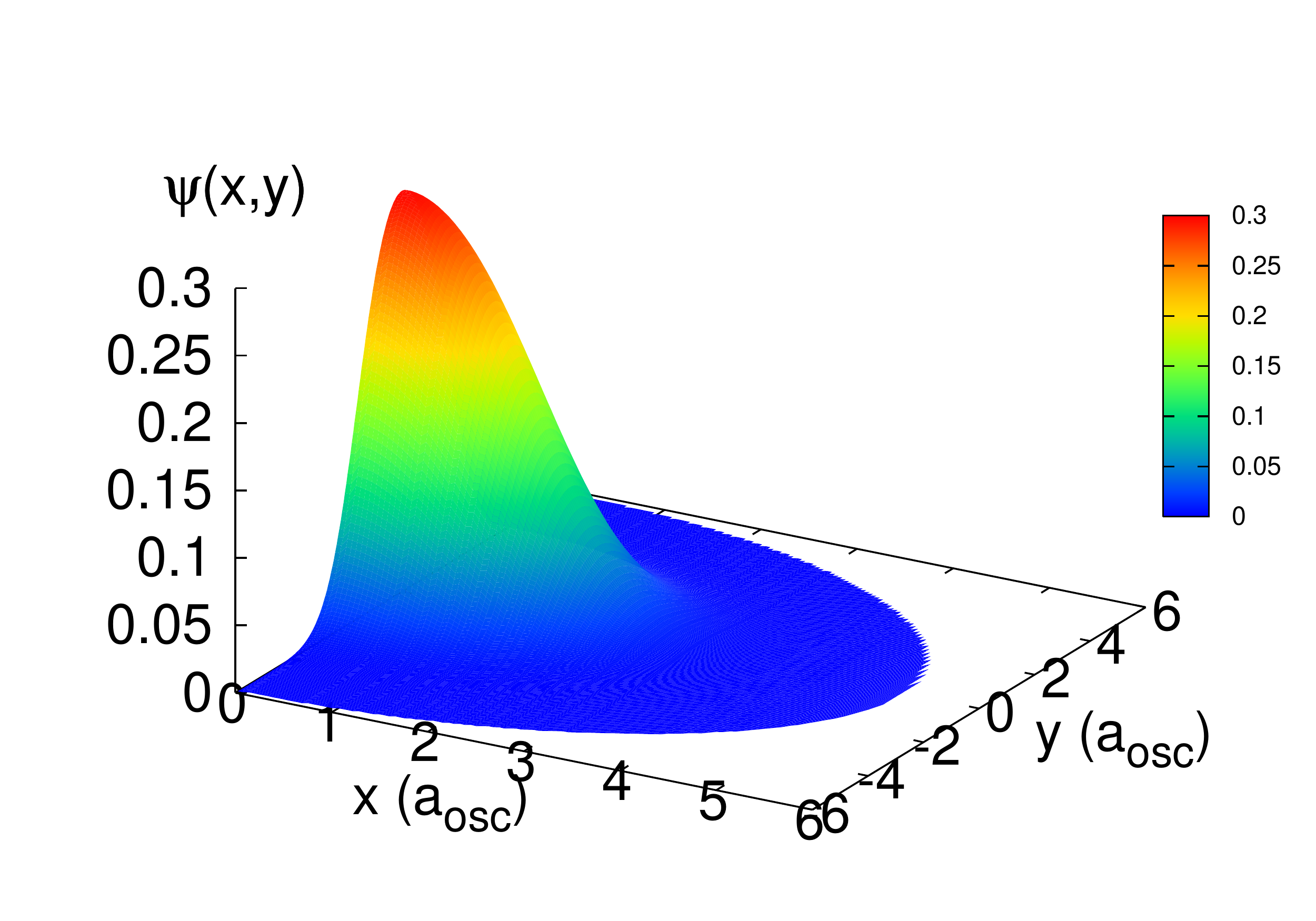}}
\end{tabular}
\caption{The semi-analytic and numerical profiles for a binary condensate
with $N_1=N_2=10,000$, $\alpha=0.8$, $a_{osc}=9.566\times10^{-7}m$,
$a_{11}=0.5a_0$, $a_{22}=1.0a_0$, and
$a_{12}=0.8a_0$. Starting from left, the upper panel shows the semi-analytic
profiles for the first and the second component, exactly below them in lower
panel are wave function profiles obtained by numerically solving
Eq.(\ref{2d.gp}).}
\label{2d_ansatz_results}
\end{figure}
\end{center}

This can now be minimized numerically to determine variational parameters and 
hence the ground state wave functions. The results of minimization with the
parameters satisfying $N_ia_{ij}/a_{osc}\sim1$ are shown in
Fig.\ref{2d_ansatz_results} along with the corresponding numerical results.

  It should be noted that the scenario of all the non-linearities to be equal 
($N_1=N_2$ and $a_{ii}=a_{ij}$) is not equivalent to single component 
condensate with $a_{ii}=a_{ij}=a_s$ as the $s$-wave scattering length and 
total number of atoms equal to $N_1+N_2$. This is due to fact that we are 
still treating the two components as two different species having order 
parameters $\phi_1(r,z)$ and $\phi_2(r,z)$.  It means that experimentally, 
even if for two different hyperfine states of an isotope above condition is 
satisfied, the system will be still a binary system having a pair of ground 
state wave functions $\phi_1(r,z)$ and $\phi_2(r,z)$ instead of single ground 
state wave function for single component BEC. Quantum mechanically, it means 
that for the system to behave as a single species BEC, the two components 
needs to be indistinguishable with the same wave function $\phi(r,z)$.  


\subsection{Quasi-1D condensates}
\label{section-2}

When the radial trapping frequency is much larger than axial trapping 
frequency ($\alpha\ll1$), and the TBEC is in the weakly interacting regime 
$a_iN_i|\psi(z)|^2\ll1$, the order parameter can be factorized into radial 
and axial parts
\begin{equation}
  \phi_i(r,z) = \xi_i(r)\psi_i(z),
\label{eq.psi}
\end{equation}
where $\xi_i(r)$ is the normalized ground state of radial trapping potential 
$V_i^{\rm rad} = r^2/2$. From Eq.(\ref{escal_axisym}), after integrating 
out the radial order parameter, the energy of the quasi-1D system is
\begin{eqnarray}
  E & = & \int_{-\infty}^{\infty}\left[\sum_{i=1}^2N_i\left(\frac{1}{2}|
         \nabla_z\psi_i|^2+ V_i^a(z)|\psi_i|^2 + 
	 \right. \right.\nonumber \\
    &   & \left. \left.N_i\frac{u_{ii}}{2}|\psi_i|^4\right) 
	 + N_1N_2u_{12}|\psi_1|^2|\psi_2|^2\right]dz, 
\label{energy}			 
\end{eqnarray}
where $u_{ii} = 2a_{ii}$, $u_{12} = 2 a_{12}$, and $V_i^a(z) = \alpha^2z^2/2 
+ 1$. We analyze the ground state of the of repulsive TBEC 
$a_{ii}, a_{ij} > 0$ trapped in quasi-1D traps in both miscible 
$a_{12}\le\sqrt{a_1a_2}$ and immiscible $a_{12}>\sqrt{a_{11}a_{22}}$ 
domains. Without loss of generality we assume $a_{22}>a_{11}$ and hence,
the first species, on account of lower repulsive mean field energy, has larger
density at the center when $N_1\approx N_2$. We consider 
\begin{eqnarray}
 \psi_1(z) & = & a e^{-(z-\gamma)^2/(2b^2)},\nonumber\\
 \psi_2(z) & = & (f + c z^2)e^{-(z+\delta)^2/(2d^2)},
 \label{variational_ansatz}
\end{eqnarray}
as our ansatz for the two  order parameters with $a,~\gamma,~b,~f,~c,~\delta$, 
and $d$ as variational parameters. Here $\gamma$ and $\delta$ represent the 
location of the order parameter maxima and center of mass motion. This is an 
apt ansatz as it describes the smooth transition between
miscible and immiscible phases of the TBEC in a very natural way.
 
 In the miscible domain, the parameter $c$ is a measure of flatness of the 
$\psi_2$ profile which arises from the larger intra-species repulsion energy. 
However, in the immiscible domain, it is the degree of separation between the
two species due to the higher inter-species repulsion energy.  The order
parameters, like in previous subsection, satisfy the normalization conditions
\begin{equation}
 \int_{-\infty}^{\infty}dz |\psi_i(z)|^2 = 1,
 \label{normalization}
\end{equation}
where $\psi_{i}(z)=   \sqrt{a_{\rm osc}/N_i}\Psi(za_{\rm osc})$, and hence
the number of independent variational parameters is reduced to five. From 
Eq.(\ref{energy}), Eq.(\ref{variational_ansatz}), and 
Eq.(\ref{normalization}), we obtain the expression for energy as a function 
of five independent parameters. It is a complicated eighth degree polynomial 
(given in appendix) and can not be solved analytically. However, one can 
treat it as a nonlinear optimization problem and use numerical schemes like 
Nelder-Meade to find a solution \cite{Press}. Like in Eq.(\ref{2d.gp}), 
the coupled 1D GP equations
\begin{equation}
  \left[ -\frac{1}{2}\frac{\partial ^2}{\partial z ^2} 
         + V_i^a(z) + 
  \sum_{j=1}^2 g_{ij}|\psi_j|^2 \right]\psi_{i} = \mu_i\psi_{i},
\label{1d.gp}
\end{equation}
are obtained when $ {\cal E} = E - \mu_iN_i$ is variationally extremized with
$\psi_i^\star$ as the variational parameters, where $g_{ii} = N_iu_{ii}$ 
and $g_{ij} = N_ju_{ij}$.


\begin{figure}[h]
\begin{center}
\includegraphics[width=9cm]{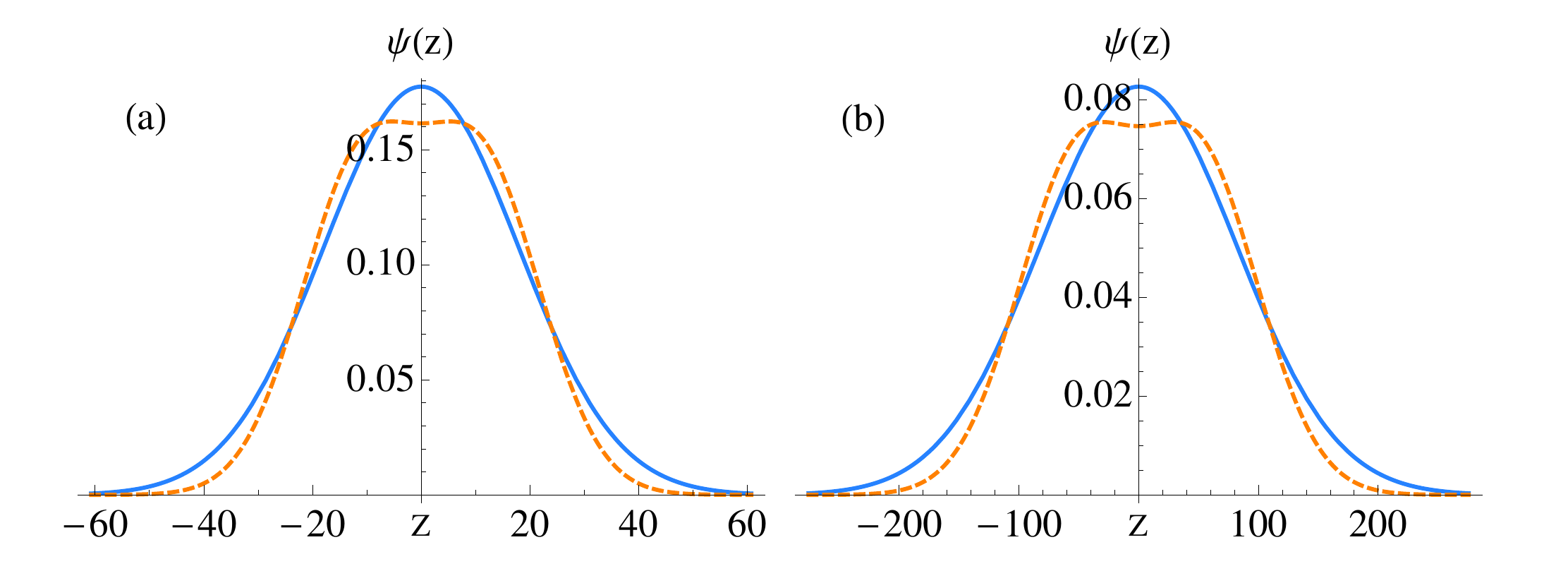}
\caption{\label{non_overlap} Order parameters $\psi_i$ of the two components
         from the variational calculations with equal intra-species 
         scattering length of $51a_0$ but no inter-species interaction 
         ($a_{ij}=0$). Solid (blue) and dashed (orange) correspond to the first
         and second species, respectively. The figure on the left hand side is 
         for $N_1=N_2=1000$, while that on the right hand side is for 
         $N_1=N_2=100,000$. The trapping potential parameters  are 
         $\alpha=0.02$ and $a_{\rm osc} = 9.566\times10^{-7}m$. For
         (a) $N_i a_{ii}/a_{osc} = 2.821$ and for (b)
         $N_i a_{ii}/a_{osc} = 282.137$. 
        }
\end{center}
\end{figure}

\subsubsection{Miscible domain with $a_{12}=0$}
\label{subsubsection-2-b-i}

 As mentioned earlier, the inequality $a_{12} < \sqrt{a_{11}a_{22}}$ defines 
the miscible domain. In this parameter range, the two order parameters  
overlap and in the limiting case of $a_{12}=0$, the two maxima coincide if 
$a_{11} = a_{22}$.  Furthermore, in the non-interacting limit 
$a_{ij}\rightarrow 0$, the profiles of the two order parameters are identical
for the same number of atoms ($ N_1=N_2$). For the special case of identical 
intra-species scattering lengths ($a_{11}=a_{22}$), finite inter-species 
interaction ($a_{12}\neq0 $), and identical number of atoms ($N_1=N_2$), an 
appropriate ansatz is 
\cite{Navarro}
\begin{eqnarray}
 \psi_1(z) & = & A e^{-(z-B)^2/(2W^2)}e^{i(C + Dz + Ez^2)},\nonumber\\
 \psi_2(z) & = & A e^{-(z-B)^2/(2W^2)}e^{i(C - Dz + Ez^2)},
 \label{complex_ansatz}
\end{eqnarray}
originally introduced to describe dynamics of coupled solitons in nonlinear
optical fibers \cite{Muraki}. Here, the parameters $A$, $B$, $C$, $D$, $E$, 
and $W$ are assumed to be time dependent and represent the amplitude, 
position, phase, wave number, chirp, and width of the Gaussian ansatz, 
respectively. In our ansatz too, a more symmetric choice of the order 
parameters
\begin{equation}
 \psi_i(z) = (f_i + c_i z^2) e^{-(z-\gamma_i)^2/(2b_i^2)},
 \label{symm_ansatz}
\end{equation}
can represent the special case mentioned here. The expression of the energy
is then much more complicated, and the situation considered being too 
restrictive, we do not consider this for further analysis and discussion.
However, to examine the ground state geometry for a varied range of 
parameters, interaction strengths, and number of atoms, the ansatz in  
Eq.(\ref{variational_ansatz}) is an ideal choice.

\begin{figure}[h]
\begin{center}
\includegraphics[width=6cm]{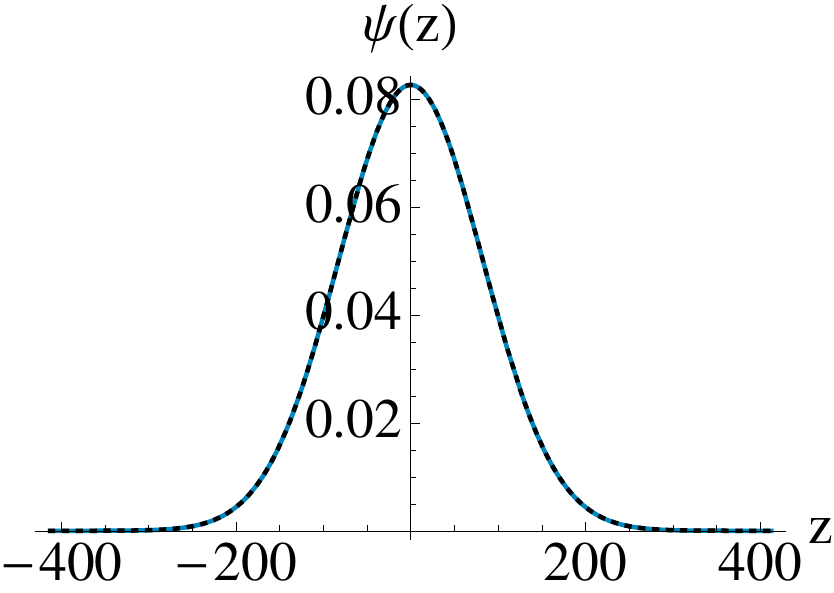}
\caption{\label{ansatz_21} Order parameters $\psi_i$ of the two components
         from the variational calculations using Eq.(\ref{complex_ansatz})
         with equal intra-species scattering length of $51a_0$
         but no inter-species interaction ($a_{ij}=0$).
         Solid (turquoise) and dotted (black) correspond to the first
         and second species, respectively. The number of atoms and
         trapping potential parameters are same as those in
         Fig.\ref{non_overlap}(b).
        }
\end{center}
\end{figure}

 With our ansatz, when $a_{11}=a_{22}\neq 0$, $a_{12}=0$,  and $N_1 = N_2$, 
the $cz^2$ dependence in $\psi_2$ accounts for the self interaction. For 
this reason, the profile of $\psi_2$ is broader and more accurate, whereas 
$\psi_1$ is a Gaussian and does not reflect the effect of the self interaction 
in the profile in an equally precise manner as $\psi_2$. To examine the 
density profiles as function of the nonlinearity, arising from the mean field 
interaction, consider TBEC with $N_i=100, 1000,$ and $10000$. As a specific 
case take $a_{11} =a_{22}=51$, which corresponds to $^{85}$Rb \cite{Cornish}, 
however, to begin with set $a_{12}=0$. The later is experimentally not 
realizable in $^{85}$Rb, but it is a good reference for a comparative study 
on the role of inter-species interactions. The order parameters for the 
different $N_i$ with the chosen parameters are as shown in 
Fig.\ref{non_overlap}. As mentioned earlier, the $\psi_2$ 
profiles are flatter and closer to the numerical values. Ideally, the two 
profiles should be identical as there is no inter-species interaction 
($a_{12}=0$). So the difference between the profiles of the two order
parameters is an indication of the error due to the Gaussian ansatz of 
$\psi_1$. From the figures, the deviation grows as $N_i$ is increased, which
is expected as the mean field contribution to $E$ is quadratic in $N$. 
In the weakly interacting domain $c<1$ and contributions to $E$ from higher 
order terms of $c$ are negligible. Retaining only the linear terms, the energy
correction arising from the mean field is
\begin{eqnarray}
 \Delta E  &\approx &\frac{cfN_2\sqrt{\pi } }{4}\left[
                \frac{2\delta ^2 - d^2}{d} + d\alpha ^2 (3 d^4 
                + 12 d^2 \delta ^2  \right .
               \nonumber \\
           &   & + 4\delta ^4) + \sqrt{2}df^2 g_{22}(d^2+4 \delta ^2)\bigg ].
  \label{delta_e}
\end{eqnarray}
For identical trapping potentials $V_1=V_2$, the order parameters $\psi_i$
are centered at the origin ( $z=0$) and hence $\delta\approx 0$. The energy
correction then simplifies to 
\begin{equation}
 \Delta E \approx \frac{cdfN_2\sqrt{\pi } }{4}(
                3\alpha ^2 d^4 + \sqrt{2}d^2f^2 g_{22}  - 1).
  \label{delta_e_coincide}
\end{equation}
The three terms are the leading order corrections from the trapping potential,
mean field, and kinetic energy, respectively. Considering that in quasi 1D 
case $\alpha\ll1$, the correction arising from $c$ is a competition between 
the mean field and kinetic energy contributions; at lower values of $g_{22}$, 
the kinetic energy correction dominates and $\Delta E$ is negative. 
As mentioned earlier for  identical intra-species scattering lengths
($a_{11}=a_{22}$), Eq.(\ref{complex_ansatz}) is more appropriate ansatz
and for parameters of Fig.\ref{non_overlap}(b) the ground state geometry
of the TBEC is shown in Fig.\ref{ansatz_21}


\subsubsection{Miscible domain with $a_{12}>0$}
\label{subsubsection-2-b-ii}

\begin{figure}[h]
\begin{center}
\includegraphics[width=8.5cm]{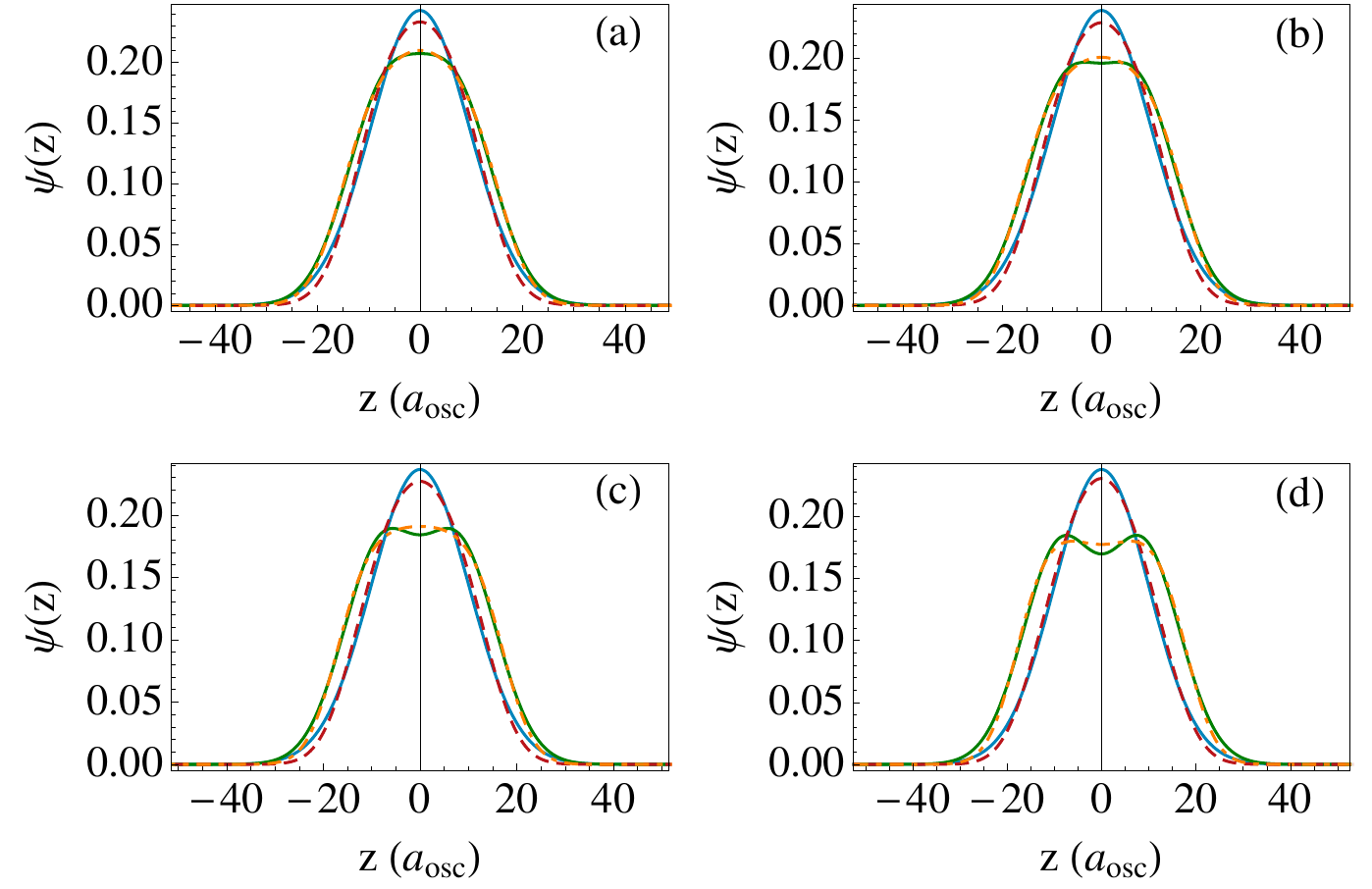}
\caption{\label{miscible_profilesp}
        Ground state wave functions for the two components of the TBEC 
        with $N_1 = N_2 = 10000$, $g_{11} = 0.553$, $g_{22} = 2g_{11}$, 
        $\alpha=0.02$ and $a_{osc} = 9.566\times10^{-7}m$. Starting from 
        left, the values of $g_{12}$ are $0.1g_{22}, 0.3g_{22}$ in upper 
        panel; and $0.5g_{22}$ and $0.7g_{22}$ in lower panel, respectively.
        The criterion, $N_j a_{ij}/a_{osc}=g_{ij}/2\sim1$, for the
        applicability of the variational ansatz is satisfied for each case. 
        Turquoise (reddish-brown) and dark-green(orange) curves correspond to 
        first and second component, respectively; solid curves are
        semi-analytic results, while dashed ones are the numerical solutions
        of Eq.(\ref{1d.gp}).
        }
\end{center}
\end{figure}

 Finite inter-species interaction ($a_{12}\neq 0 $) modifies the
density profiles of the two species in a dramatic way. As $a_{12}$ is 
increased, the species with the higher repulsion energy, second in the 
present case, is repelled from the center of trap. This is noticeable in the 
profiles of TBEC shown in Fig.(\ref{miscible_profilesp}). The lower density
of the second species, around the center of the trap decreases inter-species
density product $|\psi_1|^2|\psi_2|^2$ and minimizes the total energy. However,
this must be in proportion with the higher energy from trapping potential due 
to broader density profile. For identical trapping potentials ($V_1=V_2$), the 
profiles of the two species are centered at the origin ($ \gamma=\delta=0$) and 
symmetric. From Eq.(\ref{q1d_e12}) in appendix, the inter-species interaction 
energy is 
\begin{eqnarray}
  E_{12}^{\rm sym} & = &\frac{\sqrt{\pi}a^2bdN_1 g_{12} } {4(b^2 
              + d^2)^{9/2}}\bigg[ 4d^8f^2 + 4b^2d^6(4f^2 + cd^2f)
                   \nonumber            \\
         &   & + b^8(4f^2 + 4cd^2f + 3c^2d^4) + 2 b^6 d^2(8f^2 + 3c^2d^4 
                   \nonumber            \\
         &   & + 6cd^2f) + b^4d^4 (24f^2 + 3c^2d^4 + 12cd^2f) \bigg].
\label{q1d_misc_e12}
\end{eqnarray}
Geometrically, with higher $a_{12}$ the shape of $\psi_2$ near potential 
minima undergoes a smooth transition from Gaussian to flat 
Fig. \ref{miscible_profilesp}(a-b) and then, to parabola with a local minima 
at the center of the trap Fig. \ref{miscible_profilesp}(c-d).
When $\psi_2$ is flat around the center, to a very good approximation 
$\psi_2'(z) =\psi_2''(z)=0$ for $z<1$. With our ansatz $\psi_2''(0)$ is not
zero when $a_{12} > 0$, so there are deviations from the actual solutions. 
Deviations are negligible when $0 < a_{12}\ll 1$, however, these grow
and become quite prominent when $\psi_2$ is constant around the center of 
the trap.

At the center of the trap $\psi_1$ has the global maximum and $\psi_1'(0) = 0$.
Close to the center the order parameter
\begin{equation}
   \psi_1 (z) \approx  \psi_1(0) + \frac{z^2}{2}\psi_1''(0),
\end{equation}
when $z<1$. Gaussian ansatz of $\psi_1$ defined in 
Eq.(\ref{variational_ansatz}) then gives
\begin{equation}
   \psi_1 (z) \approx  a - \frac{z^2}{2} \frac{a}{b^2}.
\end{equation}
For the parameter domain where $\psi_2$ is constant close to the center
for $ z<1$, we can write
\begin{equation}
   \psi_2 (z) \approx  \psi_2(0).
\end{equation}
After substituting these expressions of $\psi_i$ in Eq.(\ref{1d.gp}), the
chemical potential of the two species obtained from the zeroth order equations
are
\begin{eqnarray}
  \mu_1 & = & \frac{1}{2b^2} + g_{11}a^2 + g_{12}\psi_2(0) + 1, \nonumber \\
  \mu_2 & = & g_{22}\psi_2^2(0)+ g_{21}a^2 + 1.
\end{eqnarray} 
Since $\psi_i'(0)=0$, the first order terms are zero, and hence second order 
equation of $\psi_1$ is
\begin{equation}
  \left [ \frac{\alpha^2}{2} + \frac{3g_{11}}{2}\psi_1(0)\psi_1''(0)\right ] 
  \psi_1(0) + \frac{g_{12}}{2}\psi_2(0)\psi_1''(0) = \frac{\mu_1}{2}
  \psi_1''(0).
\end{equation}
Further simplifications provide the condition
\begin{equation}
      g_{11} = \frac{1}{2a^2}\left (\frac{1}{2b^2} + \alpha^2b^2 + 1  \right )
\end{equation}
on the intra-species interaction of the first species. Similarly, the second 
order equation of $\psi_2$ is 
\begin{equation}
  \left [ \frac{\alpha^2}{2} + g_{21}\psi_1(0)\psi_1''(0)\right ]\psi_2(0) = 0.
\end{equation}
This defines the inter-species interaction as
\begin{equation}
      g_{21} = \frac{\alpha^2b^2}{2a^2},
\end{equation}
to obtain a constant profile of $\psi_2$ around center of the trap. In other
words when this condition holds true, the repulsive inter-species interaction
balances the effect of confining potential. And the net outcome is an 
effective potential around the center which is constant. The $\psi_2$ profile 
acquires a local minimum at the center when  $g_{21} > \alpha^2b^2/(2a^2)$, 
our ansatz is then the appropriate form of $\psi_2$.


\subsubsection{Immiscible domain: symmetric profile}
\label{subsubsection-2-b-iii}

 Phase separation occurs when the TBEC is immiscible. The density profiles 
of the two species are then spatially separated.  Condition for immiscibility
is $a_{12}>\sqrt{a_{11}a_{22}}$ when the TBEC is confined within a square 
well potential \cite{Ao}. In which case, the trapping potential is flat and 
except for the surface effects, limited to within healing length
$\xi=\sqrt{\hbar ^2/(2m\mu)}$ from the boundary, there are no trapping 
potential induced density variations. The TBEC is then homogeneous in the 
miscible domain and phase separated in the immiscible domain, where the 
interface acquires a geometry which minimizes the surface energy. Here, the 
density distributions in the bulk is entirely dependent on the strength of 
interactions. And transition into phase separated domain is well defined. 

  With harmonic trapping potentials the densities, to begin with, are not 
homogeneous in the miscible phase. Even when $ a_{12}<\sqrt{a_{11}a_{22}}$, 
miscible domain discussed previously, there is a separation between the two 
maxima of the densities. When $ N_1\approx N_2$ and $a_{11}\approx a_{22}$, the
density profiles are symmetric about the center ($\delta=\gamma=0 $) and 
total energy of the TBEC is
\begin{eqnarray}
   E & = & \frac{\sqrt{\pi}}{4}\left[\frac{a^2 N_1}{b} + 
           a^2 b^3\alpha ^2N_1 + \frac{N_2}{4d}(4f^2 - 4cd^2f
                        \right.\nonumber\\
     &   & + 7c^2d^4)+ \frac{d\alpha ^2 N_2}{4}( 4d^2f^2 + 12 cd^4f + 
           15c^2d^6)                          \nonumber\\
     &   & + \sqrt{2} a^4 b N_1 g_{11} + \sqrt{2} d \left ( f^4 + cd^2f^3 
           + \frac{9}{8} c^2d^4f^2 
                        \right.  \nonumber\\
     &   & \left.\left. + \frac{15}{16}c^3d^6f + \frac{105}{256}c^4d^8\right ) 
           N_2 g_{22}  \right ] +  E_{12}^{\rm sym},
\label{imm_sym}
\end{eqnarray}
where $E_{12}^{\rm sym} $ is the inter-species interaction energy defined
in Eq.(\ref{q1d_misc_e12}). For the TBEC in square well described earlier, the 
criterion for phase separation \cite{Ao} is the emergence of a dip in the 
total density $\rho_1(z) + \rho_2(z) = |\psi_1(z)|^2 + |\psi_2(z)|^2$
at the interface. Such a criterion does not apply for TBEC in harmonic 
potentials. Instead we define phase separation as the state when the 
maxima of the densities are well separated. To quantify, define 
$z_{\rm e}$ as the point where the densities of the two species are equal, 
which implies
\begin{equation}
  ae^{-z_{\rm e}^2/(2b^2)} = (f + cz_{\rm e}^2)e^{-z_{\rm e}^2/(2d^2)}.
\end{equation}
For large $a_{12}$, the order parameter $\psi_2(z)$ is bimodal. The two 
maxima are solutions of $\psi_2'(z)=0$ and located at
\begin{equation}
  z_{\rm m}= \pm \left ( \frac{4cd^2 - f}{c} \right )^{1/2}.
\end{equation}
The maxima of $\psi_2$ are well separated from $\psi_1$ when 
$z_{\rm m} > z_{\rm e}$. We define this as the criterion for phase separation
of TBEC in harmonic potentials. This definition of phase separation is an 
appropriate one in very weakly interacting TBECs $Na/a_{osc}\approx1$, but 
not so in TF regime $Na/a_{osc}\gg1$, where $f=0$ is a better criterion 
to define phase separation.

\begin{figure}[h]
\begin{center}
\includegraphics[width=8.5cm]{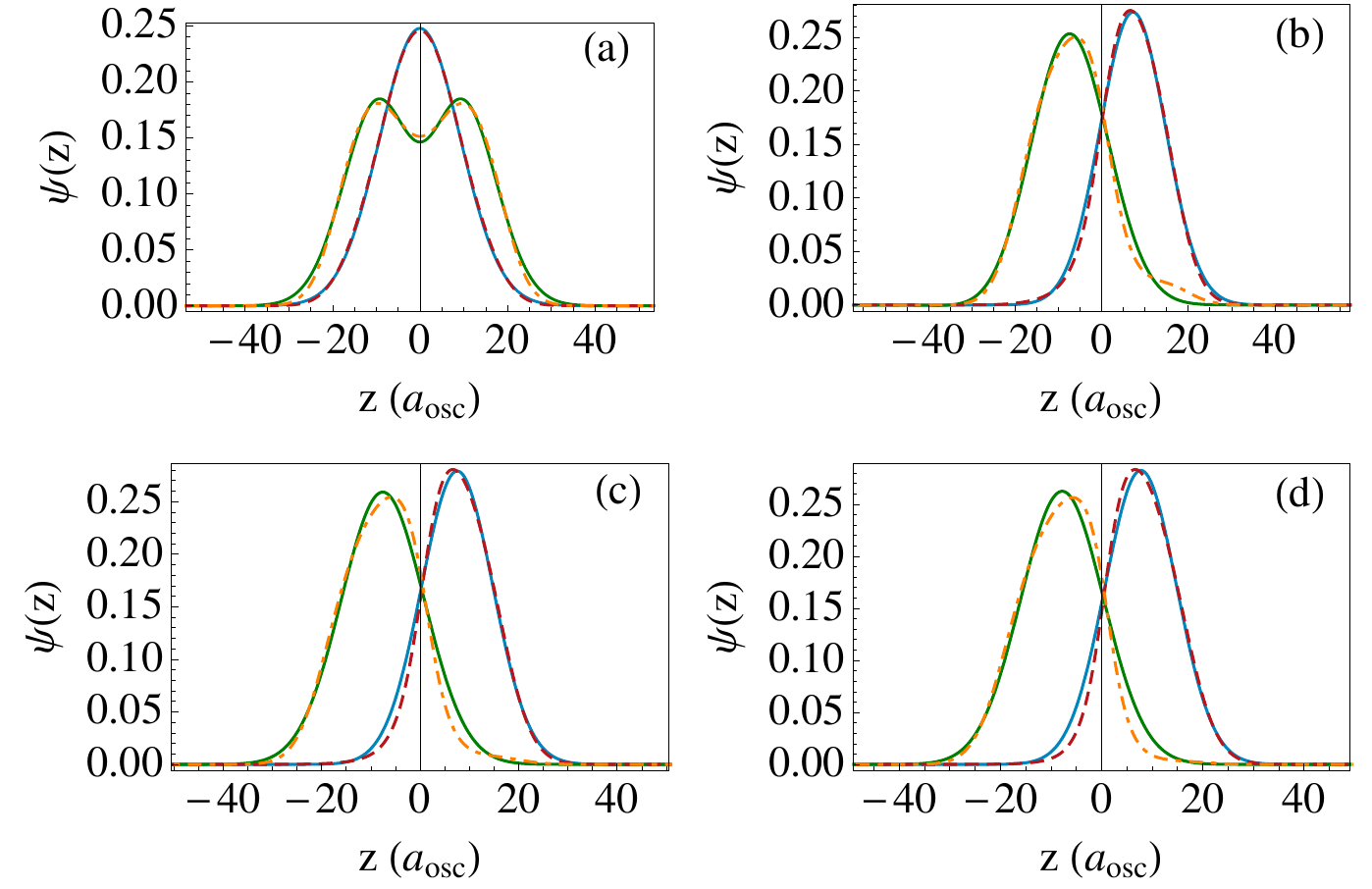}
\caption{\label{imiscible_profiles}
        Ground state wave functions for the two components of the TBEC with
        $N_1 = N_2 = 10000$, $g_{11} = 0.553$, $g_{22} = 2g_{11}$, 
        $\alpha=0.02$ and $a_{osc} = 9.566\times10^{-7}m$. Starting from 
        left, the values of $g_{12}$ are $0.9g_{22},1.2g_{22}$ in upper panel; 
        and $1.4g_{22}$ and $1.6g_{22}$ in lower panel,respectively. The 
        criterion, $N_j a_{ij}/a_{osc}=g_{ij}/2\sim1$, for the applicability 
        of the variational ansatz is satisfied for each case.Turquoise
        (reddish-brown) and dark-green (orange) curves correspond to first 
        and second component, respectively; solid curves are
        semi-analytic results, while dashed ones are the numerical solutions
        of Eq.(\ref{1d.gp}).
        }
\end{center}
\end{figure}


\subsubsection{Imiscible domain: non-symmetric profile}
\label{subsubsection-2-b-iv}

 In imiscible domain, if one or both the intra-species 
non-linearities $g_{ii}$ are sufficiently small, the stationary state 
geometry with symmetric $\psi_i$ can have higher energy than the asymmetric 
stationary state with the two components lying side by side. There are two
reasons for the emergence of asymmetric ground state geometries. First, this 
geometry has only one interface region, whereas the corresponding symmetric
profile has two; as a result the interface energy is lower. And 
second, smaller kinetic energy as $\psi_i'(z)$ is prominent at the interface
and edges of the condensates.

  With our ansatz, for the asymmetric profile $c\sim0$, but $\delta$ and 
$\gamma$ are nonzero. The total energy of the TBEC is then 
\begin{eqnarray}
   E & = & \frac{\sqrt{\pi } }{4}\bigg [ \frac{a^2 N_1}{b} + 
           \frac{f^2 N_2}{d}+a^2 b\alpha ^2 N_1 (b^2+2 \gamma ^2)
                        \nonumber     \\
     &   & + df^2 \alpha ^2 N_2(d^2+2 \delta ^2) + \sqrt{2} a^4 b N_1 g_{11}
           + \sqrt{2} df^4 N_2 g_{22} \nonumber    \\
     &   & +\frac{4a^2 b d e^{-(\gamma +\delta )^2/(b^2 + d^2)}}{(b^2 + 
            d^2)^{9/2}}  (d^8 f^2 + 4b^2d^6f^2 + b^8 f^2      \nonumber  \\
     &   & + 4b^6 d^2f^2+ 6b^4 d^4f^2) N_1 g_{12}\bigg ].
\end{eqnarray}
It must be emphasized that, with our present ansatz the asymmetric density 
profiles emerges very naturally as a function of $a_{12}$. That is not the 
case with TF based semi-analytic methods which can not account for the 
existence of of asymmetric ground state geometries 
\cite{Trippenbach}.

Typical ground state geometries of the TBECs obtained by this variational 
scheme are shown in Fig.(\ref{imiscible_profiles}) along with the respective 
non-linearities and trapping potential parameters mentioned in the caption of 
each figure. We observe that the present variational scheme also explains the 
existence of the asymmetric states, as is shown in 
Fig.(\ref{imiscible_profiles}), as the ground 
state geometry. 


\subsection{Gram-Charlier expansion of $|\psi_i|^2$}
\label{subsection-2-c}

 As mentioned earlier, for the ideal case when all the non-linearities are 
small and equal, $c$ is $\approx0$. In this limit,
both the species have Gaussian distribution. In order to quantify the 
departure, brought about by interactions among the bosons, of the density 
distributions from the normalized Gaussian, we resort to an analysis 
based on Gram-Charlier series A. It is the expansion of a probability density 
function \cite{Kendall-77} in terms of the normal distribution. It can be 
used to analyze the departure from the normal distribution. If $F(z)$ is a 
nearly normal distribution with cumulants $\kappa_r$, then $F(z)$ can be 
expressed as a series consisting of product of Hermite polynomials $H_i(z)$ 
and $\kappa_r$. The truncated expression for Gram-Charlier A series up to 
fourth cumulant is
\begin{eqnarray}
  F(\bar{z}) & = &\frac{e^{-\bar{z}^2/2}}
                {\sqrt{2\pi\sigma^2}}\left[1+
               \frac{\kappa_3}{3!\sigma^3}H_3(\bar{z})+\frac{\kappa_4}
               {4!\sigma^4}H_4(\bar{z})\right],\nonumber\\
   & = &\frac{e^{-\frac{\bar{z}^2}{2}}} {24 \sigma^5 \sqrt{2 \pi }}
        \bigg[24 \sigma^4+4 \sigma \bar{z} \left(\bar{z}^2 - 3\right) 
        \kappa _3+\nonumber\\
   &   & \left(\bar{z}^4 -6 \bar{z}^2+ 3\right) \kappa _4\bigg]
\end{eqnarray}
where $\bar{z}=(z-\mu)/\sigma$, with $\mu$ and $\sigma$ as the mean and 
standard deviation of normal distribution. The departure of a distribution 
from normal are measured in terms of skewness ( $\gamma_1$ ) and kurtosis
( $\gamma_2$ ). These quantify the asymmetry and peakedness of the 
distribution function, and are defined as
\begin{equation}
  \gamma_1 = \frac{\kappa_3}{\kappa_2^{3/2}},
  ~~\gamma_2 = \frac{\kappa_4}{\kappa_2^2},
\end{equation}
where second cumulant $\kappa_2$ is equal to the variance $\sigma^2$.
The $\mu$ and $\sigma^2$ of the normal distribution are calculated
from the numerically calculated $\psi_i(z)$ using the definitions
\begin{eqnarray}
   \mu & = & \int z|\psi(z)|^2dz,\nonumber\\
   \sigma^2 & =  & \int(z-\mu)^2|\psi(z)|^2dz. \nonumber
\end{eqnarray}
\begin{figure}[h]
\begin{center}
\includegraphics[width=8.5cm]{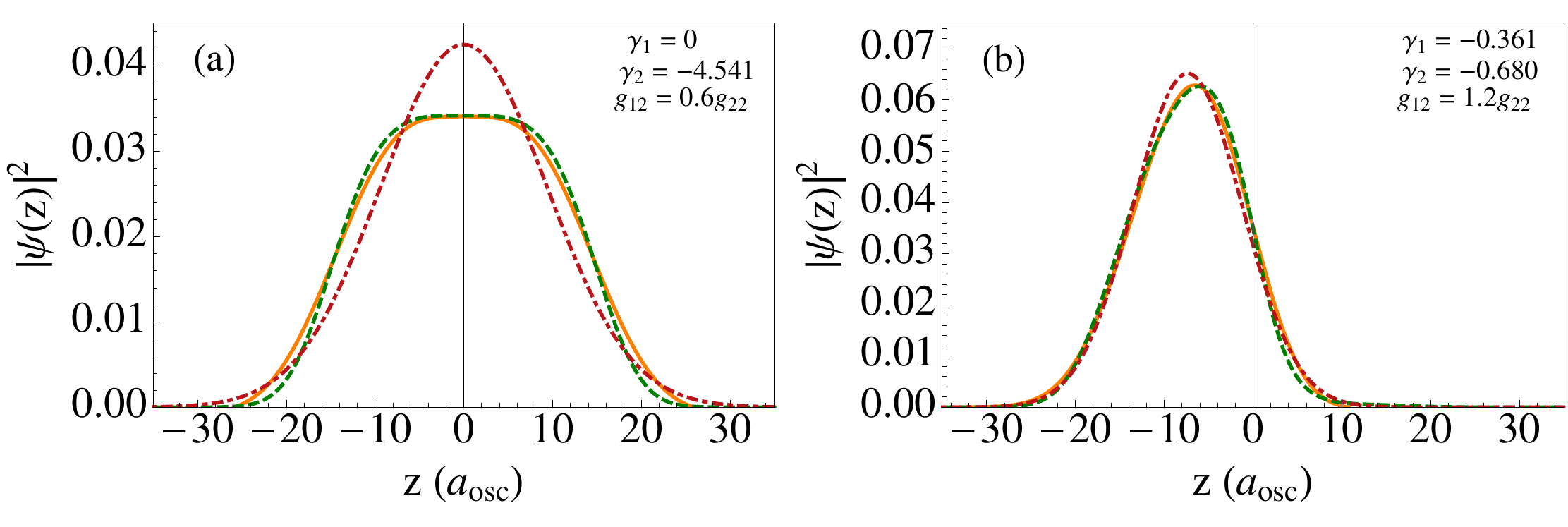}
\caption{\label{gram_charlier_profiles}
     The solid orange curve is the approximate $|\psi_2|^2 $
     of the TBEC with $N_1=N_2=10,000,~a_{\text{osc}}=
     9.566\times10^{-7},~g_{11}= 0.553,~\text{and}~g_{22} = 2g_{11}$, 
     obtained by fitting Gram-Charlier A series over numerically obtained 
     density profile (dashed dark-green curve). In (a)$g_{12} = 0.6g_{22}$, 
     while in (b) $g_{12} = 1.2g_{22}$. The dot-dashed  reddish brown shows 
     the density profile of normal distribution with respect to which 
     $\gamma_1$ and $\gamma_2$ of numerical density profile have been 
     calculated. For first figure, the variational parameters
     $f$, $c$, $\delta$, and $d$ are $0.177,~0.001~,0~,~8.675$, 
     respectively; while Gram-Charlier values are $0.185,~0,~0,13.2836$, 
     respectively. For second figure, semi-analytic values of $f$, $c$, 
     $\delta$, and $d$ are $0.267,~0,~8.529,~9.567$,respectively;
     while Gram-Charlier values are $0.270,~0,~7.358,~8.662$, respectively. 
     }
\end{center}
\end{figure}
For example, Fig.(\ref{gram_charlier_profiles}) shows the results of 
Gram-Charlier analysis of density profile for the second component of 
the TBEC in miscible and immiscible domains. As is evident from the figure, 
in the miscible domain, density profile (solid-orange curve) is symmetric and 
relatively flatter (platykurtic) compared to normal distribution 
(dot-dashed reddish-brown curve) about origin, and consequently it has zero 
skewness and negative kurtosis; whereas in second figure, the density 
distribution has a longer right tail (quantified by positive skewness) and 
narrow distribution about the mean (leptokurtic) as compared to normal 
distribution. From Gram-Charlier analysis, we can extract the coefficient of 
$z^4$ (excluding exponential part) and compare it with semi-analytic values 
of $c^2$. For both the cases in Fig.(\ref{gram_charlier_profiles}), the two 
values are same up to second decimal place.


\section{TBECs in optical lattices}

In this section, we analyze the ground state geometry of TBECs in optical 
lattices with the ansatz we have introduced. For this, we consider the 
optical lattice potential, generated by a pair of orthogonally polarized 
counter propagating laser beams along axial direction, in presence of the 
axisymmetric harmonic trapping potential. The period of the lattice 
potential is half of the laser wavelength. The net external potential 
experienced by TBEC (in scaled units) is the sum total of the two 
potentials,
\begin{equation}
   V_i(r,z) = \frac{1}{2}(r^2 + \alpha_i^2 z^2)+
	            V_0\cos^2\left(\frac{2\pi z}{\lambda}\right),
\label{optical_lat_pot}							
\end{equation}
where $V_0=sE_r$ is the depth of potential well at each lattice site and
$\lambda$ is the wavelength of the laser. Here, $E_r=(2\pi/\lambda)^2/2$ is
the recoil energy of laser light photon and $s$ is the lattice depth scaling 
parameter. As in the previous section,
in weakly interacting regime, the TBEC in cigar shaped traps is
like a quasi-1D system and its energy is given by Eq.(\ref{energy}).
The axial trapping potential is 
\begin{equation}
  V_i^{a}(z) = \frac{\alpha^2z^2}{2} +
               V_0\cos^2\left (\frac{2\pi z}{\lambda} \right )
\end{equation}

In tight binding approximation \cite{Trombettoni}
\begin{equation}
   \psi_i(z) = \sqrt{N_i}\sum_n\zeta_i(n)\phi_n(z),
\end{equation}
where $n$ is index of lattice sites, $\phi_n(z)
=\phi(z-n\lambda/2-\lambda/4)$ is the wave function with amplitude $\zeta(n)$ 
localized at $n$th lattice site. Using the tight binding ansatz, energy 
functional of a TBEC in optical lattice is
\begin{eqnarray}
E  & = &\sum_n\left\{\sum_{i=1}^2\bigg [-2K N_i\zeta_i(n)\zeta_i(n+1)+
        \epsilon(n)N_i|\zeta_i(n)|^2+  \right.       \nonumber   \\
   &   &\left.\left. \frac{\Lambda_{ii}}{2}N_i|\zeta_i(n)|^4 \right ] +
        \Lambda_{12}N_1|\zeta_1(n)|^2|\zeta_2(n)|^2\right\},
\label{energy-ol}			 
\end{eqnarray}
where $\Lambda_{ii},~\Lambda_{12},~K$ and $\epsilon(n)$ are defined as
\begin{eqnarray}
   \Lambda_{ii} & = & 2 a_{ii}N_i\int\phi_n^4 dz,    \nonumber    \\
   \Lambda_{12} & = & 2 a_{12}N_2\int\phi_n^4 dz,    \nonumber    \\
   K & = & -\int \left[ \frac{1}{2}\nabla_z\phi_n\cdot\nabla_z\phi_{n+1}
           +\phi_nV_i^{a}(z)\phi_{n+1}\right]dz,\nonumber\\
   \epsilon(n) & = &\int \left[ \frac{1}{2}(\nabla_z\phi_n)^2+
           V_i^{a}(z)\phi_{n}^2\right]dz.
\label{K&epsilon}					 
\end{eqnarray}
To examine the stationary state of the system we consider  $\zeta_i(n)$ is 
real while deriving above relation.


\subsection{Ground state}

In the weakly interacting regime, the localized wave function $\phi_n$ can be 
taken as ground state wave function of the lattice site,
\begin{equation}
   \phi_n(z) = \frac{1}{\pi^{1/4}\sqrt{\sigma}}e^{-(z-n\lambda/2 -
   \lambda/4)/(2\sigma^2)},
\label{phi}
\end{equation}
where $\sigma = \sqrt{\lambda }/(8\pi^2V_0)^{1/4}$. Ideally, the energy 
associated with each individual lattice sites can be minimized to evaluate 
$\phi_n(z)$.  Neglecting the energy due to harmonic trapping potential and 
interaction energy in comparison to potential energy due to optical lattice 
potential, the energy of the $n$th lattice site is approximately
\begin{equation}
  \int\left \{ \frac{1}{2}\left(\frac{d\phi_n}{dz}\right)^2 + V_0
  \left(\frac{2\pi}{\lambda}\right)^2\left[x - \frac{\lambda}{4} (2n + 1)
  \right]^2 \right \}dz,
\end{equation}
where we have used $V_0\cos^2(2\pi z/\lambda)\approx V_0(2\pi/\lambda)^2
(z-n\lambda/2-\lambda/4)^2$ as the potential at the $n$th lattice site.
The expressions for $\epsilon(n)$ and $K$ are
\begin{eqnarray}
   \epsilon(n) & = & \frac{V_0}{2}\left (1 -e^{-\frac{4 \pi ^2 \sigma ^2}
                     {\lambda ^2}} \right ) + \frac{1}{32\sigma ^2}\left[8 
                     + 8\alpha ^2 \sigma ^4\right . \nonumber \\
               &   & \left. + \alpha ^2\sigma ^2 \lambda ^2(1  + 
                     2n)^2 \right],  \\
K &= &-e^{-\frac{\lambda ^2}{16 \sigma ^2}} \bigg \{ \frac{V_0}{2}\left (1 + 
          e^{-\frac{4 \pi ^2 \sigma ^2} {\lambda ^2}} \right ) +
      \frac{1}{32\sigma ^4} \left[4 \sigma ^4\lambda ^2 \alpha ^2
            \right.   \nonumber \\
  &  & \left. \times (1 + n)^2 - \lambda ^2 + 8 (\sigma ^2 +
       \alpha ^2 \sigma ^6)\right]\bigg \}.
\end{eqnarray}
Similarly, the energy per boson $\bar E = E/(N_1 + N_2)$ can be calculated, 
however, as to be expected the expression is complicated and long.
Following our ansatz, we consider
\begin{eqnarray}
   \zeta_1(n) & = & a e^{-(n-\gamma)^2/(2b^2)},\nonumber\\
   \zeta_2(n) & = & (f + cn^2)e^{-(n+\delta)^2/(2d^2)},
\label{envelope_profiles} 
\end{eqnarray}
as the envelope profiles with five independent parameters. The length scale 
associated with the lattice potential, in scaled units, is 
$a_{\rm latt} = \sqrt{\omega/\omega_l}$, where $\omega_l$ is the laser 
frequency of the optical lattice. Considering that the frequency of the
harmonic trapping potential $\omega$ is at the most Khz and $\omega_l$ is in 
the optical region $ a_{\rm latt}\ll 1$. Hence, there are large number of 
lattice sites within the envelope profile. This implies that $b\gg\sigma$
and $d \gg \sigma$, where $\sigma$ is the width of the lattice ground state
$\phi_n$. We can replace summation over lattice sites with integration 
over the same variable in Eq.(\ref{energy-ol}). In fact, due
to very weak trapping along axial direction, $b$ and $d$ are of the order
of a few $a_{\rm osc}$ even for very weak non-linearities. It implies
that $e^{-\pi^2b^2}\rightarrow0$ and $e^{-\pi^2d^2}\rightarrow0$, and in
this limit the error incurred in replacing summation by integration approaches 
zero.  From Eq.(\ref{energy-ol}), Eq.(\ref{K&epsilon}), and 
Eq.(\ref{envelope_profiles}), we calculate the energy per particle which can be
minimized to determine the variational parameters. The typical variational
results for equal number of atoms of two species are shown in 
Fig.(\ref{phase_sep_profile_OL}).

\begin{figure}
\begin{center}
\includegraphics[width=9cm]{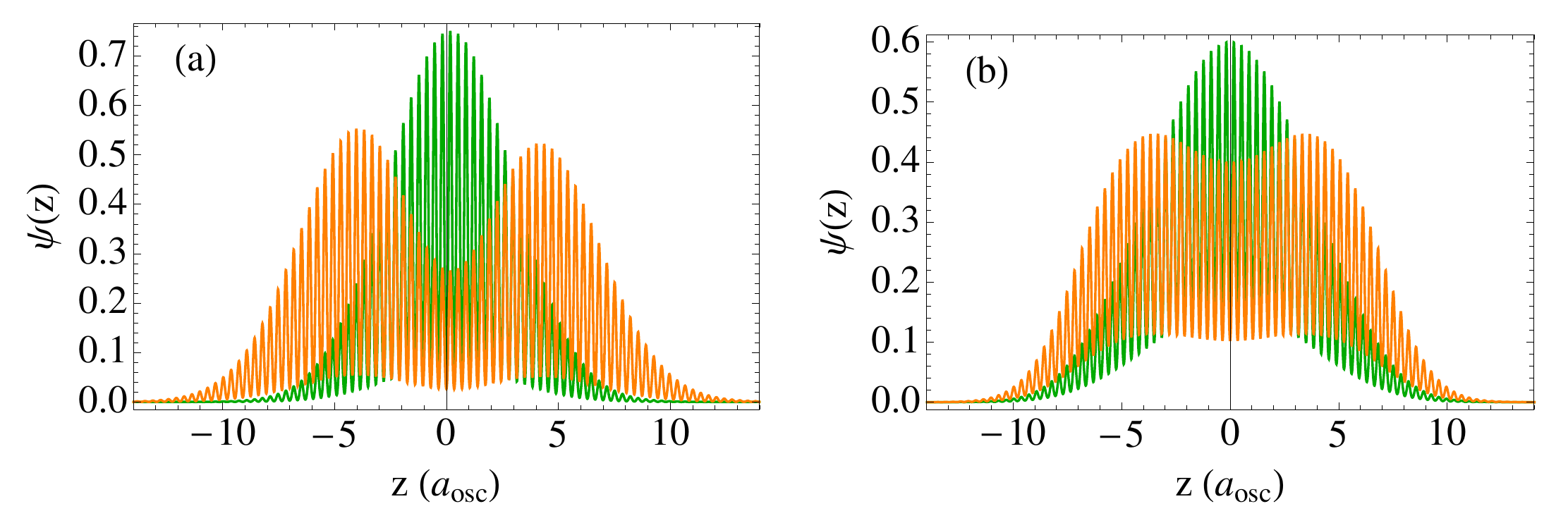}
\caption{\label{OL_symm}The stationary state solution of the TBEC with  
         $N_1 = N_2=10,000$, $g_{11}  = 0.465$, $g_{22} = 2g_{11}$,
         $g_{12}=g_{21} = 0.8g_{22}$, $\alpha=9/92$, 
         $a_{\rm osc} = 1.137\times10^{-6}m$, $\tilde{\lambda}=0.7$, and 
         $s=6.0$ as the non-linearity and trapping potential parameters, 
         respectively. The criterion, $N_j a_{ij}/a_{osc}=g_{ij}/2\sim1$,
         for the applicability of the variational ansatz is clearly
         satisfied. The left and right figures are, respectively, the 
         profiles obtained by variational minimization and numerically solving 
         quasi-1D GP equation.}
\end{center}
\end{figure}

\begin{figure}
\begin{center}
\includegraphics[width=9cm]{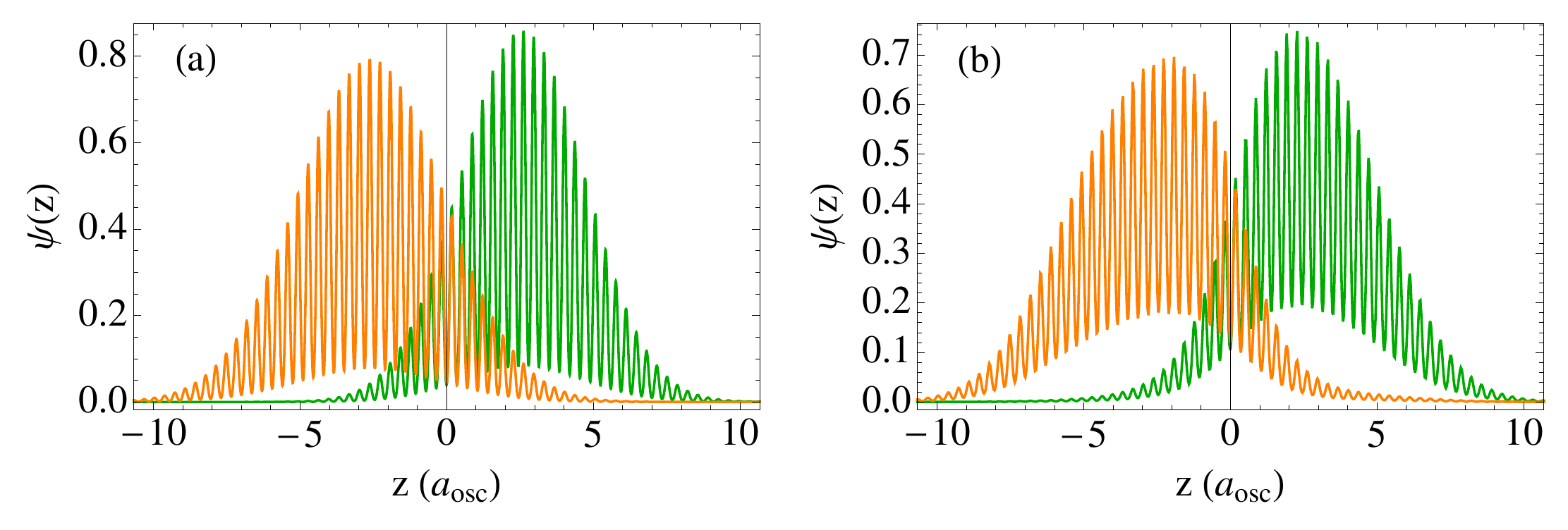}
\caption{\label{OL_asymm}The stationary state solution of the TBEC with  
         $N_1 = N_2=5,000$, $g_{11}  = 0.233$, $g_{22} = 2g_{11}$, 
         $g_{12}= g_{21} =  2g_{22}$, $\alpha=9/92$, 
         $a_{\rm osc} = 1.137\times10^{-6}m$, $\tilde{\lambda}=0.7$,
         and $s=6.0$ as the non-linearity and trapping potential parameters, 
         respectively. The criterion, $N_j a_{ij}/a_{osc}=g_{ij}/2\sim1$,
         for the applicability of the variational ansatz is clearly
         satisfied. The left and right figures are, respectively, the 
         profiles obtained by variational minimization and numerically 
         solving quasi-1D GP equation.}
\end{center}
\end{figure}

\subsubsection*{Symmetric and Non-symmetric Profiles}
Similar to TBECs in harmonic trapping potentials, ground sate geometry
of binary condensates in optical lattices can be symmetric or asymmetric. 
In miscible and weakly segregated domain 
( coherence length of the first component is smaller than the penetration 
depth of the second component \cite{Ao} ), the ground state geometries are 
symmetric. For example, a typical symmetric ground state geometry of TBEC in 
quasi-1D optical lattice is shown in Fig.(\ref{OL_symm}).

 In strongly segregated domain ( coherence length of the first
component is greater than the penetration depth of the second component), 
the ground state geometries can be asymmetric. A typical asymmetric ground 
state geometry of TBEC in quasi-1D optical lattice is shown in
 Fig.(\ref{OL_asymm}).


\subsection{Profile narrowing with lattice depth}

The density profiles of the two components in TBEC is more compact with 
deeper lattice potential.  To analyze this, we consider an ideal Bose gas
( non-interacting ) trapped in a triple 1D potential well superimposed with 
a weak harmonic potential 
\begin{equation}
  V(x)=\left \{ \begin{aligned}
          &\frac{1}{2}\alpha^2x^2 + \frac{1}{2} \beta ^2(x + d)^2  
                  && \text{if } -\frac{3 d}{2}\leqslant x<-\frac{d}{2}, \\
          &\frac{1}{2}\alpha^2x^2 + \frac{1}{2}\beta ^2x^2   
                  &&\text{if }  -\frac{d}{2}\leqslant x\leqslant \frac{d}{2}, \\
          &\frac{1}{2}\alpha^2x^2 + \frac{1}{2}\beta ^2 (x - d)^2  
                  &&\text{if }  \frac{d}{2}<x\leqslant \frac{3 d}{2}, \\
          &\infty      &&\text{if }   |x| > 3d/2,
       \end{aligned} \right .
\label{triple_pot}		
\end{equation}
where $\alpha$ and  $\beta$ are the parameters of harmonic and periodic 
potentials, respectively, and $d$ is the spatial extent of each well. This is 
most basic model to examine the nearest neighbour effects in a lattice. 
The central well, in particular, which has two nearest neighbours is a 
representation of lattice sites in optical lattices. In the experiment
set up used by Cataliotti et. al. \cite{Cataliotti}, the trap geometry 
is quasi-1D with $\omega_z/\omega_r=9/92$. For further analysis the 
equations are scaled in terms of the transverse oscillator length of the 
harmonic potential realized in aforementioned experimental set up, 
which is equivalent to setting $\alpha=9/92$. Accordingly, $\beta$ and
all the coordinates, from here after, are in scaled units. The potential 
considered is a simplified model and discontinuities are present at the 
boundary of two neighbouring wells. However, the ground state 
energy is much lower than the barrier height and it describes the underlying 
physics very well. In the tight-binding approximation, the total wave 
function is
\begin{equation}
  \psi(x) = \sum_{i=-1}^1a_{i}\phi_{i}(x), 
\end{equation}
where $i=-1,0, \text{and } 1$ represent the left, central, and right well, 
respectively.  Here the normalization is
\begin{equation}
  \int_\infty^\infty |\psi(x)|^2 dx = 1.
\end{equation}
We consider $\beta > \alpha$, which follows from the tight binding 
approximation. The width of each of these localized wave functions are 
calculated by minimizing the localized energies of each well
\begin{equation}
  E_i  = \int N_i\left[ \frac{1}{2}|\nabla \phi_i|^2 + 
         V_i(x)|\phi_i|^2 \right] dx ,
\end{equation}
where, from the previous definition
\begin{equation}
  V_i(x) = \frac{1}{2} x^2 +\frac{1}{2}\beta ^ 2 (id + x)^2.
\end{equation}
The range of each well, as defined earlier, is
$(2i-1)d/2\leqslant x \leqslant (2i+1)d/2$. Neglecting the nearest neighbour 
overlaps of wave the functions, the integration limits can be considered as 
$-\infty$ to $+\infty$. The energies are then
\begin{eqnarray}
  E_0 & = & \frac{1}{2} \eta ,  \nonumber\\
  E_{\pm 1} & = & E_0 +\delta E
  \label{E_wells} 
\end{eqnarray}
where $\eta=\sqrt{\alpha^2 + \beta ^2} $ and 
\begin{equation}
\delta E = \frac{\alpha^2 d^2}{2},
\end{equation} 
is the energy difference between side and central wells.
\begin{figure}[h]
 \includegraphics[width=8.5cm]{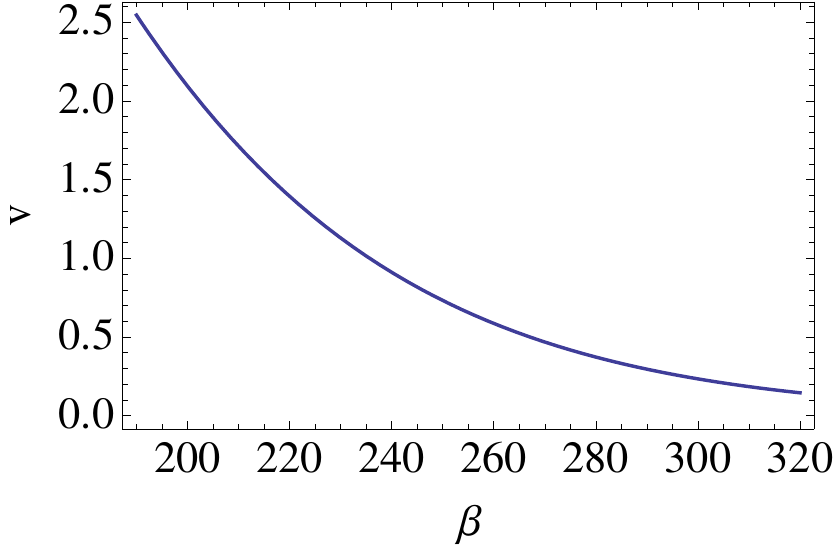}
 \caption{\label{v_beta} Variation of $v$ as a function of $\beta$.
 }
\end{figure}

\begin{figure}[h]
 \includegraphics[width=8.5cm]{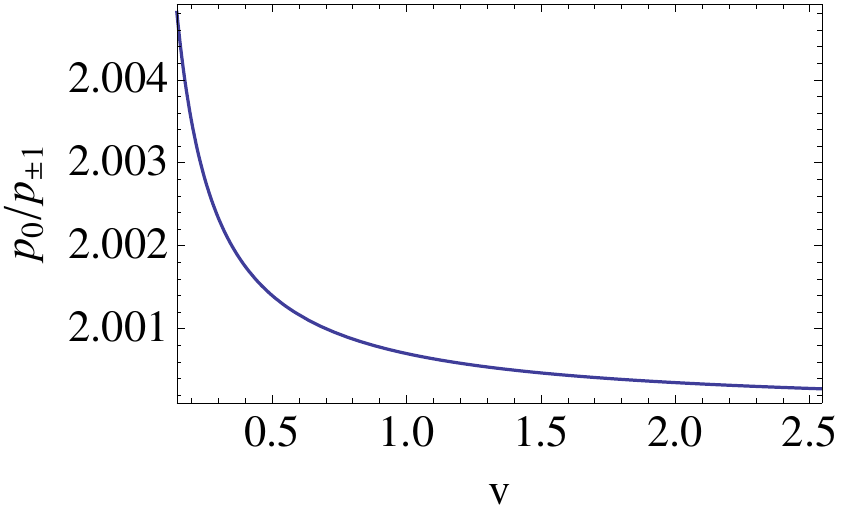}
 \caption{\label{p0_p1_ratio} The plot shows the $p_0/p_{\pm 1}$
          as a function of $v$ for the values of $\beta$ ranging 
          from $197.82$ to $312.79$. In terms of the ratio of barrier 
          height to recoil energy of laser light photon $s$ used in 
          Ref.\cite{Cataliotti}, the above mentioned range of 
          $\beta$ values is equivalent to $s$ ranging from $6$ to $15$.
         }
 \end{figure}

  For the combined system of the three wells, in tight binding approximation, 
we only consider tunneling between adjacent wells. The lowest eigen energy is 
the chemical potential of the system and the eigen vector correspond to the 
probability amplitudes for the occupancy of each well. As defined earlier,
$E_{-1}$, $E_0$, and $E_1$ are the energies localized in left, central, and 
right well, respectively. Define the tunneling matrix element between left and 
central well as
\begin{equation}
  v = \int \left[ \frac{1}{2}\nabla\phi_{-1}\cdot\nabla\phi_0
           +\phi_{-1}V(x)\phi_0\right]dx,
\end{equation}
and similarly, for the central and right wells as 
\begin{equation}
  w = \int \left[ \frac{1}{2}\nabla\phi_0\cdot\nabla\phi_1
           +\phi_0V(x)\phi_1\right]dx.
\end{equation}
For the symmetric case considered here $ v = w$, and after evaluating the 
integrals
\begin{eqnarray}
v & = & \frac{e^{-\frac{17}{4} d^2\eta}}{16 \sqrt{\pi\eta}}  
        \bigg \{ 4 d \left[ \beta ^2\left(1-2 e^{3 d^2\eta} 
        - 2 e^{4 d^2\eta}\right) - \left(1+2 e^{3 d^2\eta}\right)\right]
             \nonumber\\
  &   & -\sqrt{\pi\eta}e^{4 d^2\eta}  \left[2\eta + d^2 (5 \beta ^2 - 3)\right] 
        \text{Erf}\left(d \sqrt{\eta}\right)  
                    \nonumber  \\
  &   & - \sqrt{\pi\eta } e^{4 d^2 \eta}\left[2\eta-d^2 (11\beta ^2 + 3)\right]
        \text{Erf}\left(2 d\sqrt{\eta}\right) \bigg \}
\end{eqnarray}
In the above expression $\text{Erf}(\cdots)$ represents error function.
The Hamiltonian matrix of the system is then 
\begin{equation}
 H = \begin{pmatrix}
            E_{-1} & v   & 0 \\
            v      & E_0 & v \\
            0      & v   & E_1 
     \end{pmatrix}.
\label{h_matrix}
\end{equation}
The ground state of the system is the lowest energy eigen vector obtained 
from diagonalizing the Hamiltonian matrix. This is equivalent to solving the 
secular equation and amounts to calculating the roots of a cubic polynomial, 
which is possible analytically. The eigen values, in increasing 
order of magnitude, of the Hamiltonian in Eq.(\ref{h_matrix}) are
\begin{equation}
\frac{1}{2} (E_0+E_1-E_{12}),~~E_1,~~\frac{1}{2} (E_0+
E_1+E_{12});
\end{equation}
where $E_{12}=\sqrt{{\delta E}^2+8 v^2}$. The eigen vector of the lowest eigen 
value is
\begin{eqnarray}
   a_{\pm 1} & = & \frac{\sqrt{2} v}{\sqrt{8 v^2+\delta E \left(\delta E+E_{12}
\right)}},\nonumber\\
   a_0       & = & -\frac{\sqrt{8 v^2+\delta E
\left(\delta E+E_{12}\right)}}{\sqrt{2} 
\sqrt{{\delta E}^2+8 v^2}}, \nonumber
\end{eqnarray}
and the ratio of the probability amplitude of occupancy of central to side 
well is
\begin{equation}
  \frac{|a_0|}{|a_{\pm 1}|} = \frac{\delta E (\delta E+
  E_{12})+8 v^2}{2 v \sqrt{{\delta E}^2+8 v^2}}.
\label{ratio_a0_a1}
\end{equation}
This energy difference $\delta E$ is quite small for small values of $d$. For
example based on Ref.\cite{Cataliotti}, consider $d=0.35$, and 
$\beta=197.824$. The associated energy difference and tunneling matrix 
element are $\delta E = 0.00049494$ and $v = 2.18549$, respectively. 
Hence, for $\delta E\ll v$, using Taylor series expansion 
Eq.(\ref{ratio_a0_a1}) simplifies to
\begin{equation}
\frac{|a_0|}{|a_{\pm 1}|} \approx \sqrt{2}+\frac{\delta E }{2v}.
\label{a0_a1_ratio}
\end{equation}
Since $\delta E$ is independent of $\beta$ and $v$ decreases as $\beta$ is 
increased (see Fig.(\ref{v_beta})), the above ratio is larger with 
higher values of $\beta$. Using Eq.(\ref{a0_a1_ratio}), 
Fig.(\ref{p0_p1_ratio}) shows the variation of ratio of probability of 
occupancy of central to side wells with respect to $v$.
This relation is not valid for very large values of $\beta$ when 
$\delta E\ll v$ does not hold true.
For trapping potentials without the harmonic component ($\alpha=0$), purely 
lattice potential, the three eigenvalues are identical when the hopping
across adjacent wells is zero ($v = w = 0 $). This solution correspond to
all the atoms confined to one of the potential wells. At this stage for 
further reference we define the occupancy probability of the $i$th well as
\begin{equation}
  p_i = \left ( \int_{-\infty}^\infty \phi_i^*\psi dx\right)^2  = |a_i|^2.
\end{equation}
When $v$ and $w$ are non-zero, then the ground state has nonzero occupancies 
for all the three wells.The central well has probability of occupancy  equal 
to half ($p_0=0.5$) while the wells at the wings have probability equal to 
one fourth each ($p_{\pm 1}=0.25$). More importantly, these values 
are are independent of $\beta$. That is, no squeezing of the density profile
occurs without the harmonic potential ($\alpha=0$). 

\begin{figure}
\begin{center}
\includegraphics[width=9cm]{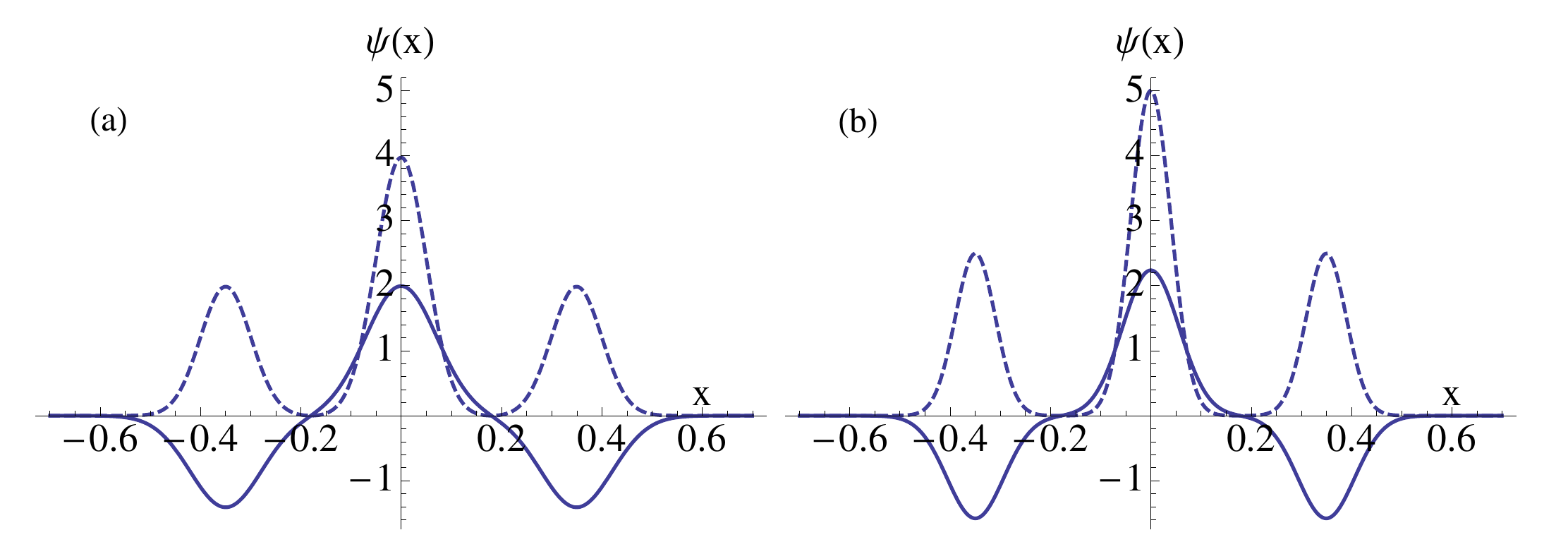}
\caption{\label{triple_well}The plots show the ground state wave function 
        $\psi(x)$ (solid curve) and probability density $|\psi(x)|^2$
         (dashed curve) for potential considered in Eq.(\ref{triple_pot})
         For figure on left side, $\alpha=9/92,d=0.35$ and 
         $\beta=197.82$ ($s=6$), while for figure on right side,
         $\alpha=9/92,d=0.35$ and $\beta=312.79$ ($s=15$)}
\label{phase_sep_profile_OL}
\end{center}
\end{figure}

 In the presence of a weak harmonic potential, there is an increase in 
$p_0 $ when $\beta$ is increased. For $\beta=197.82$, equivalent to $s=6$ in 
Ref.\cite{Cataliotti}, the occupancy probability of the central well 
is $0.500047$, this is marginally larger than the $\alpha=0$ case. 
On the other hand, the wings have occupancy probability of $0.249976$, 
slightly lower probability of occupancy than $\alpha=0$ case.  
The trend continues with further increase of $\beta$, and for $\beta=1000$ 
there are almost no atoms in the potential wells at the wings. 
Thus with increase in value of $\beta$, atoms migrate from the wings to 
central well and lead to the squeezing of the condensate. The squeezing of 
the condensate profile is evident from the comparative study of 
plots in Fig(\ref{triple_well}).

In three level approximation, a triple well problem can also be analyzed by 
considering the following classical Hamiltonian \cite{Graefe}
\begin{eqnarray}
\mathcal{H} & = &\frac{1}{2}g\left[n_{-1}^2+(1-n_{-1}-n_1)^2+n_1^2\right]
+n_{-1} \epsilon_{-1} +n_1 \epsilon_1 +\nonumber\\
& &2 \sqrt{1-n_{-1}-n_1} \left( v\sqrt{n_{-1}} \cos\theta_{-1}+w
\sqrt{n_1} \cos\theta_1\right),
\end{eqnarray}
where $n_{j}=|a_{j}|^2$ and $\theta_j = \arg(a_2)-\arg(a_j)$ are canonical
conjugate variables satisfying following equations
\begin{equation}
\frac{dn_{j}}{dt} = -\frac{\partial \mathcal{H}}{\partial \theta_j} ~ 
\text{and} ~ \frac{d\theta_{j}}{dt} = \frac{\partial \mathcal{H}}
{\partial n_j},
\end{equation} 
and $\epsilon_{-1}$ and $\epsilon_1$ are the zero point energies
of potential wells at the flanks measured with respect to zero point energy of 
central well. The stationary states of the system can be calculated by
by solving 
\begin{equation}
\frac{dn_{j}}{dt} = \frac{d\theta_{j}}{dt} = 0 
\end{equation}
In the absence of the harmonic potential $\epsilon_{-1}$ and $\epsilon_1$ are 
zero and change in tunneling amplitude has no effect 
on the occupancy of three wells. But the situation is dramatically different 
with the harmonic potential, $\epsilon_{-1}=\epsilon_1\ne0$. In this case 
probability of occupancy of central well increases with decrease in $v$. 
This is experimentally realizable by increasing depth of lattice potential. 
Similarly for four potential wells, the occupancy of the two central wells 
(whose occupancies are equal) grows as we increase $\beta$ for non 
zero $\alpha$.

\section{Conclusions}
We have studied the stationary state properties of the TBECs both in
miscible and imiscible domains using a variational ansatz. The stationary
state geometries obtained using the variational ansatz are in very good
agreement with numerical results, especially for very weakly interacting
TBECs, i.e., $Na/a_{osc}$ is of the order $1$. We have also quantified the 
departure of the variational ansatz based semi-analytic results from the 
numerical ones using Gram-Charlier analysis. Besides harmonic trapping 
potentials, the present ansatz can also be used to study the TBECs in deep 
optical lattices where coupled GP equations can be mapped into coupled 
discrete non-linear Schr\"odinger equations. In optical lattices, the density 
profiles of the component species of the TBEC get squeezed with the increase 
in the height of the potential barrier between adjacent wells. 
We have explained this phenomenon using a very simple triple well potential 
model.


\begin{acknowledgements}
We thank S. A. Silotri, B. K. Mani, and S.  Chattopadhyay for very useful 
discussions. We acknowledge the help of P. Muruganandam while doing the 
numerical calculations. The numerical computations reported in the paper
were carried on the 3 TFLOPs cluster at PRL.
\end{acknowledgements}


\section{Appendix}

\begin{equation}
   E_1  = \frac{\sqrt{\pi} N_1}{4} \biggl [
          \frac{ a^2}{b} + a^2b\alpha ^ 2(b^2 + \gamma ^2) 
          + \sqrt{2} a^4 bg_{11} \biggr ]
\end{equation}

\begin{eqnarray}
  E_2 & = &\frac{\sqrt{\pi }N_2 }{1024} \biggl \{
           \frac{64}{d} \bigl[4 f^2 - 4 c f(d^2 - 2 \delta ^2) + c^2 (7 d^4 
           \bigr. \biggr . \nonumber \\
      &   & \bigl. + 20 d^2 \delta ^2 + 4 \delta ^4)\bigr] +
           64 d \bigl[4 f^2(d^2 + 2 \delta ^2) + 4cf(3 d^4  \bigr .
                           \nonumber \\
      &   & + 12 d^2 \delta ^2 + 4 \delta ^4) + c^2(15 d^6 + 
           90 d^4 \delta ^2 + 60 d^2 \delta ^4 
                           \nonumber \\
      &   &\bigl. + 8 \delta ^6)\bigr] \alpha ^2 + \sqrt{2} d \bigl [256 f^4 
           + 256cf^3(d^2 + 4 \delta ^2) \bigr.
                           \nonumber \\ 
      &   &+ 96 c^2f^2(3 d^4 + 24 d^2 \delta ^2 + 16 \delta ^4) + 16 c^3 f 
           (15 d^6         \nonumber \\
      &   &\biggl.\bigl.+ 180 d^4 \delta ^2 + 240 d^2 \delta ^4 
           + 64 \delta ^6)+ c^4(105 d^8 + 1680 d^6 \delta ^2 
                           \nonumber \\ 
      &   &+ 3360 d^4 \delta ^4 + 1792 d^2 \delta ^6 + 
           256 \delta ^8)\bigr ] g_{22} \biggr \} 
\label{q1d_e}
\end{eqnarray}

\begin{eqnarray}
 E_{12} & = &\frac{\sqrt{\pi }a^2bdN_1g_{12}e^{-\frac{(\gamma + 
             \delta )^2}{b^2 + d^2}}}{4(b^2+ d^2)^{9/2}} 
             \left \{4 d^8(f + c \gamma ^2)^2
              \right.\nonumber\\
        &   &+4 b^2 d^6 \left[4 f^2 + c f (d^2 + 4\gamma ^2 -
             4\gamma\delta ) + c^2 \gamma ^2(3 d^2 \right.
                        \nonumber \\
        &   &\left. - 4\gamma\delta )\right]+ b^8 \left[4 f^2 + 
             4 c f (d^2 + 2 \delta ^2) + c^2 (3 d^4 + 12 d^2 \delta ^2 \right.
                        \nonumber \\
        &   &\left.  + 4 \delta ^4)\right]+ 2 b^6 d^2 \left[8 f^2 + c^2(3d^4 
             - 8\gamma\delta ^3 - 12 d^2 \delta \gamma \right.
                        \nonumber \\
        &   & \left. + 6 d^2 \delta^2 ) + 2cf(3 d^2-4 \delta  \gamma 
             + 4\delta^2 )\right] + b^4 d^4 \left[24 f^2  \right. 
                        \nonumber \\
        &   & + 3 c^2 (d^4 + 4d^2\gamma ^2 - 8d^2 \gamma\delta +
              8\gamma ^2 \delta ^2) + 4 c f (3 d^2 
                        \nonumber \\
        &   &\left. \left. + 2\gamma ^2 - 8 \gamma\delta 
             + 2\delta ^2)\right]\right \}
\label{q1d_e12}
\end{eqnarray}

\begin{eqnarray}
  c & = &\frac{2}{3d^5 + 12d^3 \delta ^2 + 4d \delta ^4}  
         \left[-d^3 f - 2df\delta ^2 + \frac{1}{\sqrt{\pi}}
         \left(3 d^5 \sqrt{\pi } \;\;\;\;\; \right. \right. 
                 \nonumber \\
    &   &\biggl. \left. - 2 d^6f^2\pi + 12d^3 \sqrt{\pi } \delta ^2
         -8 d^4f^2 \pi \delta ^2 + 4d\sqrt{\pi }\delta ^4\right)^{1/2}\biggr ]
\label{q1d_c}
\end{eqnarray}

\end{document}